\documentclass[a4paper, 12pt]{article}
\usepackage{epsfig,amssymb,euscript,xspace, color}
\usepackage{amsmath,jheppub,mathtools,empheq,amsthm}

\usepackage[T1]{fontenc} 
\usepackage{tikz,caption,subcaption,marvosym} 
\usetikzlibrary{decorations.markings,arrows,snakes}

\def\bea{\begin{eqnarray}}
\def\eea{\end{eqnarray}}
\def\be{\begin{equation}}
\def\ee{\end{equation}}

\definecolor{lightblue}{rgb}{.1,.4,.5}
\definecolor{brown1}{rgb}{.64,.43,.138}



\title{Scalar Blocks as Gravitational Wilson Networks}
\author{Atanu Bhatta, Prashanth Raman and  Nemani V Suryanarayana}
\emailAdd{batanu, prashanthr, nemani@imsc.res.in}
\affiliation{Institute of Mathematical Sciences, \\ Taramani, Chennai 600 113, India \\ \\ Homi Bhabha National Institute,  \\ Anushakti Nagar, Mumbai 400085, India}

\abstract{

In this paper we continue to develop further our prescription \href{https://arxiv.org/abs/1602.02962}{[arXiv:1602.02962]} to holographically compute the conformal partial waves of CFT correlation functions using the gravitational open Wilson network operators in the bulk. In particular, we demonstrate how to implement it to compute four-point scalar partial waves in general dimension. In the process we introduce the concept of {\it OPE modules}, that helps us simplify the computations. Our result for scalar partial waves is naturally given in terms of the Gegenbauer polynomials. We also provide a simpler proof of a previously known recursion relation for the even dimensional CFT partial waves, which naturally leads us to an odd dimensional counterpart. }

\begin{document}

\maketitle


\section{Introduction}
The correlation function of a set of primary operators in a $d$-dimensional CFT can be decomposed into its partial waves. For example, the correlation function of four scalar primary operators can be decomposed as
\bea
\langle {\cal O}_1(x_1) {\cal O}_2(x_2) {\cal O}_3(x_3) {\cal O}_4(x_4) \rangle = \sum_{{\cal O}} C_{12 {\cal O}} C^{\cal O}_{~~34} W^{(d)}_{\Delta, l} (\Delta_i, x_i)
\eea
where $C_{12 {\cal O}}$ are the OPE coefficients and the partial wave $W^{(d)}_{\Delta, l} (x_i)$ is
\bea
W^{(d)}_{\Delta, l} (\Delta_i, x_i) = \left(\tfrac{x_{24}^2}{x_{14}^2}\right)^{\tfrac{1}{2} (\Delta_1 - \Delta_2)} \left(\tfrac{x_{14}^2}{x_{13}^2}\right)^{\tfrac{1}{2} (\Delta_3 - \Delta_4)} (x_{12}^2)^{-\tfrac{1}{2}(\Delta_1+\Delta_2)} (x_{34}^2)^{-\tfrac{1}{2}(\Delta_3+\Delta_4)}  \, G_{\Delta, l} (u,v)
\eea
The pre-factor is determined by the conformal invariance and the function $G_{\Delta, l} (u,v)$ -- referred to as the conformal block -- depends only on the conformally invariant cross-ratios $u, v$. A lot is known about these conformal partial waves/blocks. For instance, a general expression for conformal partial waves (CPW) of four-point scalar correlators is given in \cite{Dolan:2011dv} (see also \cite{Dolan:2000ut, Dolan:2003hv}). Written in terms of the complex coordinates $z, \bar z$ where $u= z \, \bar z$ and $v = (1-z) \, (1- \bar z)$, closed form expressions are known for all even $d$ for scalar CPW \cite{Dolan:2011dv, SimmonsDuffin:2012uy}. Also closed form expressions for scalar conformal blocks for particular choice $z = \bar z$ are known for all dimensions \cite{ElShowk:2012ht, Hogervorst:2013kva}. Powerful recursion relations between blocks in even $d$ are found in \cite{SimmonsDuffin:2012uy}. A different choice of parametrising the cross-ratios through $z = x \, e^{i \,\theta}$ and $\bar z = x \, e^{-i \, \theta}$ was also advocated in \cite{Dolan:2011dv, Hogervorst:2013sma}.

Since AdS/CFT provides a natural avenue to answer questions in CFT$_d$ in terms of $AdS_{d+1}$ gravity (and vice versa) it is natural to ask how to compute the conformal partial waves of a given correlation function of primary operators in a CFT holographically. To achieve this two distinct prescriptions have been proposed so far in the literature:
\begin{enumerate}
\item {\bf Geodesic Witten Diagrams} \cite{Hijano:2015zsa}:  This prescription is based on the second order Einstein-Hilbert formulation of gravity in which the conformal partial waves are given by the so called geodesic Witten diagrams. This has been generalised further in \cite{Hijano:2015qja, Nishida:2016vds, Dyer:2017zef, Belavin:2017atm, Tamaoka:2017jce, Kraus:2017ezw, Anand:2017dav, Nishida:2018opl}. 

\item {\bf Gravitational Open Wilson Networks} \cite{Bhatta:2016hpz, Besken:2016ooo}: This prescription is suitable for the first order Hilbert-Palatini formulation of the bulk theory in which the conformal partial waves are given by appropriate  gravitational open Wilson networks (OWN). These are studied and generalised for 2d CFTs in \cite{Fitzpatrick:2016mtp, Besken:2017fsj, Hikida:2017ehf, Hikida:2018dxe}.
\end{enumerate}
In this paper we restrict ourselves to the second prescription, and provide further computational methods for its implementation in general dimensions. Before proceeding further let us review some essential aspects of this construction (see \cite{Bhatta:2016hpz} for more details). 

In the first-order Hilbert-Palatini formulation of $AdS_{d+1}$ gravity \cite{MacDowell:1977jt, Freidel:2005ak} the basic fields are the vielbeins $e^a$ and the spin-connections $\omega^{ab}$.  They are conveniently combined into a 1-form gauge field $A$ in the adjoint of the $so(1, d+1)$ algebra as:
\bea
\label{gaugeconnection}
A = \frac{1}{l} e^a \, M_{0a} + \frac{1}{2} \omega^{ab} M_{ab}
\eea
where $\{ M_{0a}, \, M_{ab}\}$ are the generators of $so(1, d+1)$ with $a,b = 1, \cdots, d+1$. In this theory we consider a set of gauge covariant Wilson Network operators. In particular, 
\begin{itemize}
\item  One starts with an open, directed and trivalent graph (such as in Fig. \!(1) ) whose every line (internal as well as external) carries a representation label of the (Euclidean) conformal algebra $so(1,d+1)$. 

\item The representations of interest are those non-unitary infinite dimensional irreps which are obtained by appropriate Wick rotation of the corresponding UIR of the associated Lorentzian conformal algebra $so(2,d)$ of the  CFT$_d$. Such an irrep can be labeled by $(\Delta; l_1, \cdots, l_{[d/2]})$ where $\Delta$ is the conformal weight and $l_i$ label which irrep the primary transforms in, under the boundary rotation group $so(d)$.
\end{itemize}

%
\vskip .5cm
\begin{center}
	\begin{tikzpicture} [scale = .5]
	\draw [very thick] (0,0) circle (5 cm);
	\fill [gray!15] (0,0) circle (5 cm);
	\draw [>= stealth, ->] (-4,0) -- (-3.33,0);
	\draw (-3.33,0) -- (-2.83,0);
	\draw [>= stealth, ->] (-4.93,0.93) -- (-4.43,0.43);
	\draw (-4.43,0.43) -- (-4,0);
	\filldraw [black] (-4.93,0.93) circle (2pt);
	\draw [>= stealth, ->] (-4.93,-0.93) -- (-4.43,-0.43);
	\draw (-4.43,-0.43) -- (-4,0);
	\filldraw [black] (-4.93,-0.93) circle (2pt);
	\draw (-4,0) circle (4pt);
	\draw [>= stealth, ->] (-2.83,0) -- (-2.83,0.465);
	\draw (-2.83,0.465)--(-2.83,0.93);
	\draw [>= stealth, ->] (-2.83,0) -- (-2.83,-0.465);
	\draw (-2.83,-0.465) -- (-2.83,-0.93);
	\draw (-2.83,0) circle (4pt);
	\draw [>= stealth, ->] (2.83,0) -- (3.415,0);
	\draw (4,0) -- (3.415,0);
	\draw [>= stealth, ->] (4,0) -- (4.465,0.465);
	\draw (4.93,0.93) -- (4,0);
	\filldraw [black] (4.93,0.93) circle (2pt);
	\draw [>= stealth, ->] (4,0) -- (4.465,-0.465);
	\draw (4.93,-0.93) -- (4.465,-0.465);
	\filldraw [black] (4.93,-0.93) circle (2pt);
	\draw (4,0) circle (4pt);
	\draw [>= stealth, ->] (2.83, 0.93) -- (2.83,0.465);
	\draw (2.83, 0.465) -- (2.83,-2.4);
	\draw [>= stealth, ->] (2.83,-2.4) -- (3.615,-2.4);
	\draw (3.615,-2.4) -- (4.4,-2.4);
	\draw (2.83,0) circle (4pt);
	\filldraw [black] (4.4,-2.4) circle (2pt);
	\draw (0,4) -- (0,3.45);
	\draw [>= stealth, ->]  (0,2.83) -- (0,3.45);
	\draw (0.93,4.93) -- (0.43,4.43);
	\draw [>= stealth, ->] (0,4) -- (0.43, 4.43);
	\filldraw [black] (0.93,4.93) circle (2pt);
	\draw (-0.93,4.93) -- (-0.43,4.43);
	\draw [>= stealth, ->] (0,4)--(-0.43,4.43) ;
	\filldraw [black] (-0.93,4.93) circle (2pt);
	\draw (0,4) circle (4pt);
	\draw [>= stealth, ->] (-0.93,2.83) -- (-0.43,2.83);
	\draw [>= stealth, ->] (-0.43,2.83) -- (0.43,2.83);
	\draw (0.43,2.83) -- (0.93,2.83);
	\draw (0,2.83) circle (4pt);
	\draw [>= stealth, ->] (0,-2.83) -- (0,-3.415);
	\draw (0,-4) -- (0,-3.415);
	\draw [>= stealth, ->] (0,-4) -- (-0.465,-4.465);
	\draw (0.93,-4.93) -- (0.465,-4.465);
	\draw [>= stealth, ->] (0,-4) -- (0.465,-4.465);
	\filldraw [black] (0.93,-4.93) circle (2pt);
	\draw (-0.93,-4.93) -- (-0.465,-4.465);
	\filldraw [black] (-0.93,-4.93) circle (2pt);
	\draw (0,-4) circle (4pt);
	\draw [>= stealth, ->] (-0.93,-2.83) -- (-0.465,-2.83);
	\draw [>= stealth, ->] (-0.465,-2.83) -- (0.465,-2.83);
	\draw (0.465,-2.83) -- (0.93,-2.83);
	\draw (0,-2.83) circle (4pt);
	\draw [>= stealth, ->] (-2.83,2.83) -- (-3.25,3.25);
	\draw (-3.54,3.54) -- (-3.25,3.25);
	\filldraw [black] (-3.54,3.54) circle (2pt);
	\draw [>= stealth, ->] (-2.83,1.83) -- (-2.83,2.33);
	\draw (-2.83,2.33) -- (-2.83,2.83);
	\draw [>= stealth, ->] (-2.83,2.83) -- (-2.33,2.83);
	\draw (-2.33,2.83) -- (-1.83,2.83);
	\draw (-2.83,2.83) circle (4pt);
	\draw [loosely dashed] (-2.83,1.83) -- (-2.83,0.93);
	\draw [loosely dashed] (-1.83,2.83) -- (-0.93,2.83);
	\draw (3.54,3.54) -- (3.185,3.185);
	\draw [>= stealth, ->] (2.83,2.83) -- (3.185,3.185);
	\filldraw [black] (3.54,3.54) circle (2pt);
	\draw [>= stealth, ->] (2.83,2.83) -- (2.83,2.33);
	\draw (2.83,2.33) -- (2.83,1.83);
	\draw [>= stealth, ->] (1.83,2.83) -- (2.33,2.83);
	\draw (2.33,2.83) -- (2.83,2.83);
	\draw (2.83,2.83) circle (4pt);
	\draw [loosely dashed] (2.83,1.83) -- (2.83,0.93);
	\draw [loosely dashed] (1.83,2.83) -- (0.93,2.83);
	\draw [>= stealth, ->] (-2.83,-2.83) -- (-3.185,-3.185);
	\draw (-3.54,-3.54) -- (-3.185,-3.185);
	\filldraw [black] (-3.54,-3.54) circle (2pt);
	\draw [>= stealth, ->] (-2.83,-1.83) -- (-2.83,-2.33);
	\draw (-2.83,-2.33) -- (-2.83,-2.83);
	\draw [>= stealth, ->] (-2.83,-2.83) -- (-2.33,-2.83);
	\draw (-2.33,-2.83) -- (-1.83,-2.83);
	\draw (-2.83,-2.83) circle (4pt);
	\draw [loosely dashed] (-2.83,-1.83) -- (-2.83,-0.93);
	\draw [loosely dashed] (-1.83,-2.83) -- (-0.93,-2.83);
	\draw [loosely dashed] (0.93,-2.83) -- (2.83,-2.83);
	\draw [>= stealth, ->] (2.83,-2.83) -- (2.83,-3.49);
	\draw (2.83,-3.49) -- (2.83,-4.15);
	\filldraw [black] (2.83,-4.15) circle (2pt) ;
	\node at (3.4,-4.7) {\tiny $ \mathcal{O}_{R_n}(x_n)$};
	\node at (5.8,-2.4) {\tiny $ \mathcal{O}_{R_1}(x_1)$};
	\node at (6.3,-0.96) {\tiny $ \mathcal{O}_{R_2}(x_2)$};
	\draw [very thick,dotted] (5.5,0.93) arc (10:20:5.5 cm);
	\draw [very thick,dotted] (1.35,-5.4) arc (-76:-85:6 cm);
	\node at (3.415,0.35) {\tiny $ R$};
	\node at (2.3,0.5) {\tiny $R'$};
	\node at (3.4,2.33) {\tiny $R''$};
	\node at (-0.7,-3.415) {\tiny $R^{(k)}$};
	\end{tikzpicture}
	
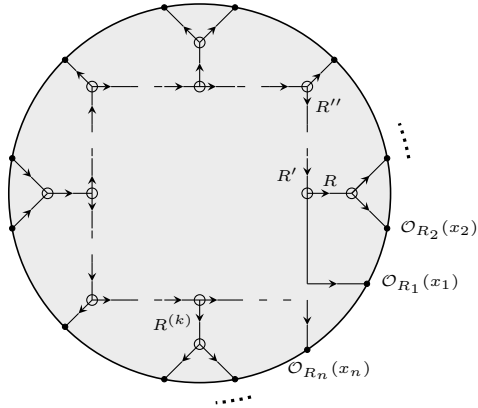
\captionof{figure}{A typical directed trivalent Open Wilson Network}
\end{center}
%
%
 %
%
%
%
\begin{itemize}

\item Next, one associates an open Wilson line (OWL) operator 
\bea
W_y^x(R, C) = P \, {\rm exp} [\int_y^x A]
\eea
for 1-form $A$ in (\ref{gaugeconnection}) to the line labelled by the irrep R, connecting the points $x$ and $y$ in the graph.

\item At every trivalent vertex where three lines carrying representation labels $(R_1, R_2, R_3)$ join -- one glues the corresponding OWLs with the appropriate Clebsch-Gordan coefficients to make the vertex gauge invariant. 

\item One projects each of the external lines onto {\it Cap States} \cite{Verlinde:2015qfa, Miyaji:2015fia, Nakayama:2015mva} -- a set of states in the conformal module $R$ labelling that leg that also provides a finite-dimensional irrep of the $so(d+1)$ subalgebra whose generators are $M_{ab}$ used in (\ref{gaugeconnection}).

\item One evaluates these OWNs for the gauge connection $A$ that corresponds to the Euclidean Poincare $AdS_{d+1}$. Such a gauge connection has to satisfy the flatness condition
\bea
F:= dA + A \wedge A =0.
\eea
\item Finally one takes the external legs to the boundary and reads out the leading component of the OWN - and these compute the relevant conformal partial waves.

\end{itemize}
This leading component of the OWN satisfies the conformal Ward identities and conformal Casimir equations expected of the partial waves of a correlator of primaries that are inserted at the points on the boundary to which the end points of the external legs of the OWN approach.

In short, the basic ingredients needed to compute our OWNs are  (i) Wilson lines, (ii) CG coefficient and (iii) the cap states. These were found for $d=2$ in \cite{Bhatta:2016hpz} for the most general case. When the external legs were taken to the boundary the computation reduced to simple Feynman-like rules that require the knowledge of what we called {\it legs} (more precisely the conformal wave functions) and the CG coefficients. The explicit computations using these rules to find the global conformal blocks of correlators of primary operators (with any conformal dimension and spin) was demonstrated explicitly for $d=2$ in \cite{Bhatta:2016hpz} (see also \cite{Besken:2016ooo}). 

Even though the general prescription for computing the partial waves of correlators of  any set of primaries (in arbitrary representations of the rotation group of the boundary theory) in general CFT$_d$ using OWNs was laid down in \cite{Bhatta:2016hpz}, the actual computations in higher dimensions could not be carried through as some of the necessary ingredients were missing. In this work we would like to report some progress in this direction. In particular, we will demonstrate how to implement our prescription explicitly for the scalar CPW $W^{(d)}_{\Delta, 0} (\Delta_i, x_i)$ in any CFT$_d$. Our results include a simplification of the computation of OWN{\it s} using the concept of {\it OPE modules} - which are close analogues of the OPE blocks that were studied in the literature \cite{Czech:2016xec, Boer:2016pqk}. With this simplification we compute the scalar 4-point blocks in general dimension and show that our prescription reproduces the known answers \cite{Dolan:2011dv}. Remarkably, our results are naturally given in Gegenbauer polynomial basis \cite{Dolan:2011dv, Hogervorst:2013sma}. Further, we show that there is a non-trivial recursion relation that emerges from our prescription which relates the scalar blocks in $d+2$ dimensions to those of $d$ dimensions. This relation reproduces the one in \cite{SimmonsDuffin:2012uy} in the context of even $d$, and provides an analogue for the odd $d$ cases. 

The rest of the paper is organised as follows: The section \ref{sec2} contains the construction of the modules and the conformal wave functions required for the computation of scalar blocks. We also introduce the concept of OPE module here and use it to carry out the computation of the 4-point scalar blocks in general dimensions. The section \ref{sec3} contains details of how our answers match with several known results in $d \le 4$. In section \ref{sec4} we derive recursion relations between different dimensions. In section \ref{sec5} we include a couple of generalisations: most general bulk analysis in $d=1$, more general bulk geometries in $d=2$. We provide a discussion of our results and open questions in section \ref{sec6}. The appendices contain some relevant mathematical results used in the text. 

\section{Scalar OWN in General Dimensions}
\label{sec2}
In this section we would like to provide details on how to explicitly compute the OWN{\it s} in $AdS_{d+1}$ spaces, with all lines (both external and internal) carrying scalar representations. 

\subsection{Collecting the Ingredients}
As has been alluded to in the introduction the basic ingredients are Wilson lines, cap states and CG coefficients. We start with collecting these ingredients first.
\vskip .5cm 
\noindent\underline{\bf Wilson Lines}
\vskip .5cm 
We will be evaluating the OWN in the background of the Euclidean $AdS_{d+1}$ geometry with ${\mathbb R}^d$ boundary ({\it i.e}, Poincare $AdS_{d+1}$) with the metric:
\bea
\label{eadsdplus1}
l^{-2} \, ds^2_{AdS_{d+1}} = d \rho^2 + e^{2 \rho} dx^i dx^i.
\eea
For this, working with the frame:
\bea
e^i = l \, e^\rho \, dx^i, ~ i = 1, \cdots, d, ~~ e^{d+1} = l \, d\rho
\eea
the Wilson line reduces to 
\bea
W_y^x(R, C) = P \, {\rm exp} [\int_y^x A] = g(x) \, g^{-1} (y)
\eea
as was shown in \cite{Bhatta:2016hpz}, with
\bea
g(x) = e^{-\rho \, M_{0,d+1}} e^{- x_a \, (M_{0,a}+M_{a,d+1})} g_0 \, ,
\eea
where the algebra generators are taken in the representation $R$ of $so(1, d+1)$.
Using the standard identification of $so(1, d+1)$ generators as the conformal generators of ${\mathbb R}^d$: 
\bea
\label{algdefs}
D=-M_{0,d+1}, ~~ P_\alpha = M_{0\alpha} + M_{\alpha,d+1}, ~~~ K_\alpha = -M_{0\alpha} + M_{\alpha,d+1}, ~~ {\rm and}~~~  M_{\alpha\beta}
\eea
where $\alpha, \beta =1, \cdots, d$, the coset element $g(x)$ reads:
\bea
\label{generaldg}
g(x) = e^{\rho \, D} e^{- x^a \, P_a} g_0.
\eea 
This gives us the Wilson lines.
\vskip .5cm 
\noindent\underline{\bf The Scalar Caps}
\vskip .5cm 
To project the external legs of the OWN operator we seek states, in the representation space $R$ carried by that external leg, that transform in a (finite dimensional) irrep of the subalgebra $so(d+1)$ with generators $\{M_{\alpha\beta}, M_{\alpha, d+1}\}$  \cite{Nakayama:2015mva}. In particular, for the scalar cap this finite dimensional representation is the trivial one, that is, annihilated by $\{M_{\alpha\beta}, M_{\alpha, d+1}\}$. Let us now construct these states.

In terms of the generators in (\ref{algdefs}) the $so(1, d+1)$ algebra reads
\bea
[M_{\alpha\beta}, P_\gamma] &=& - (\delta_{\alpha\gamma} P_\beta - \delta_{\beta\gamma} P_\alpha), ~~[M_{\alpha\beta}, K_\gamma] = - (\delta_{\alpha\gamma} K_\beta - \delta_{\beta\gamma} K_\alpha), \cr
 [P_\alpha, K_\beta] &=& -2 \, M_{\alpha\beta} -2 \, \delta_{\alpha\beta} \, D, ~~ [D, P_\alpha] = P_\alpha, ~~ [D, K_\alpha] = -K_\alpha \, , \cr
 [M_{\alpha\beta}, M_{\gamma\delta}] &=& \delta_{\alpha \delta} M_{\beta \gamma} + \delta_{\beta \gamma} M_{\alpha\delta} - \delta_{\alpha \gamma} M_{\beta\delta} - \delta_{\beta\delta} M_{\alpha\gamma}\, .
\eea
We work with irreps $R$ of $so(1, d+1)$ that become UIR of $so(2,d)$ obtained by a Wick rotation. This implies the following reality conditions
\bea
M_{0,d+1}^\dagger = M_{0,d+1}, ~~ M_{0\alpha}^\dagger = - M_{0,\alpha}, ~~ M_{\alpha, d+1}^\dagger = M_{\alpha, d+1}, ~~ M_{\alpha\beta}^\dagger = - M_{\alpha\beta} \, .
\eea
In terms of the generators in (\ref{algdefs}) these mean:
\bea
\label{reality}
D^\dagger = D, ~~ P_\alpha^\dagger = K_\alpha, ~~ M_{\alpha\beta}^\dagger = - M_{\alpha\beta}.
\eea
Then the scalar cap state $|\Delta \rangle\!\rangle$ is defined to be a state in the scalar module $(\Delta, l_i =0)$ that satisfies the conditions:
\bea
M_{\alpha\beta} |\Delta \rangle\!\rangle =  (P_\alpha + K_\alpha) |\Delta \rangle\!\rangle = 0.
\eea
We can construct it as a linear combination of states in the module over the scalar primary (lowest weight) state $|\Delta \rangle$ which satisfies 
\bea
\label{scalarcapeqn}
D |\Delta \rangle = \Delta \, |\Delta \rangle, ~~ M_{\alpha\beta} \, |\Delta \rangle = K_\alpha \, |\Delta \rangle =0.
\eea
Rest of the basis states of the module take the form $|\Delta, k_i \rangle = {\cal N}_{\vec k} \, P_1^{k_1} \cdots P_d^{k_d} |\Delta \rangle$. The solution to the scalar cap state equation (\ref{scalarcapeqn}) was provided first in \cite{Nakayama:2015mva} (see also \cite{ Verlinde:2015qfa, Miyaji:2015fia} for $d=2$ case). We rederive it here for completeness. For this note that the cap state has to be a singlet under $so(d)$ and therefore can only depend on $P_\alpha P^\alpha$. So write
\bea
\label{eq:scalar_cap}
|\Delta \rangle\!\rangle = \sum_{n=0}^\infty C_n (\Delta, d) \, (P_\alpha P^\alpha)^n |\Delta \rangle \, ,
\eea
and impose $(P_\alpha + K_\alpha) |\Delta \rangle\!\rangle = 0$ to determine the coefficients $C_n$. Carrying out this straightforward exercise gives
\bea
\label{scalarcapcns}
C_n (\Delta, d) = \frac{(-1)^n}{2^{2n} n! \, \left(\Delta - \mu \right)_n}
\eea
With these (\ref{eq:scalar_cap}) can be seen to be equivalent to the one in \cite{Nakayama:2015mva} using the definition of the Bessel function of first kind $J_\alpha (x)$. We will need the dual (conjugate under (\ref{reality})) of this cap state which is given by:
\bea
\label{hcscalarcap}
\langle \!\langle \Delta | = \sum_{n=0}^\infty C_n (\Delta, d) \, \langle \Delta | \, (K_\alpha K^\alpha)^n
\eea
with the same $C_n$ as in (\ref{scalarcapcns}).\footnote{This scalar cap in the $d=2$ case can be seen to be equivalent to that with $h=\bar h$ cap used in \cite{Bhatta:2016hpz} (see also \cite{Miyaji:2015fia} and more recently \cite{Castro:2018srf} for a different perspective).}

In fact one can obtain more general cap states. For instance, in the case of $d=2$, we \cite{Bhatta:2016hpz} provided expressions for cap states in the module over the primary state $|h, \bar h \rangle$ that transform under $(j,m)$ representation of $so(3)$ algebra. In other dimensions one should seek caps that transform under arbitrary finite dimensional irreps of $so(d+1)$ -- to be used in computing the OWN{\it s} with primaries that are not just scalars (see (\ref{vectorcap}) for the vector cap state -- provided for illustration). We however will not pursue this further here.
\vskip .5cm
\noindent\underline{\bf CG coefficients}
\vskip .5cm
The last ingredient in the computation of the OWN expectation values is the Clebsch-Gordan coefficients (CGC) of the gauge algebra $so(1,d+1)$. Some of these are known -- see for instance \cite{kerimov1984}. Those are however not in a form that lends itself readily to our purposes. So here we propose a method to derive them using the 3-point functions. 

For this first recall that the CG coefficients are defined as the invariant tensors in the product of three representations. That is, the CGC that appear in the tensor product decomposition $R_1 \otimes R_2  \rightarrow R_3$ satisfy:
\bea
\label{cgcinv}
R_1[g(x)]_{{\bf m}_1 {\bf m}_1'} R_2[g(x)]_{{\bf m}_2 {\bf m}_2'} C^{R_1, R_2; R_3}_{{\bf m}_1' ,{\bf m}_2'; {\bf m}_3'} R_3[g(x)^{-1}]_{{\bf m}_3' {\bf m}_3} = C^{R_1, R_2; R_3}_{{\bf m}_1 ,{\bf m}_2; {\bf m}_3} 
\eea
where $R_i[g(x)]_{{\bf m}_i {\bf m}_i'}$ is used to denote the matrix elements of $g(x)$ in the representation $R_i$, whose basis elements are collectively labelled  by ${\bf m}_i$. In terms of the algebra elements $M_{AB}$ with $A, B = 0, 1, \cdots, d+1$, this eq.\eqref{cgcinv} reads:
\bea
\label{cgcrecursion}
R_1[M_{AB}]_{{\bf m}_1 {\bf m}_1'} C^{R_1, R_2; R_3}_{{\bf m}_1' ,{\bf m}_2; {\bf m}_3} +  R_2[M_{AB}]_{{\bf m}_2 {\bf m}_2'} C^{R_1, R_2; R_3}_{{\bf m}_1 ,{\bf m}_2'; {\bf m}_3} =   C^{R_1, R_2; R_3}_{{\bf m}_1 ,{\bf m}_2; {\bf m}_3'}  R_3[M_{AB}]_{{\bf m}_3' {\bf m}_3}
\eea
which is the recursion relation that determines the CGC. Now we argue that this is equivalent to the conformal Ward identity of the 3-point function of primary operators corresponding to the irreps $(R_1, R_2, R_3)$. The prescription of \cite{Bhatta:2016hpz} for the 3-point function of scalar primaries is to extract the leading term, {\it i.e,} the coefficient of $e^{-\rho (\Delta_1 + \Delta_2 + \Delta_3)}$ term -- in the boundary limit of
\bea
\label{own3ptfn}
\langle\! \langle \Delta_1| g(x_1) |\Delta_1, {\bf m}_1 \rangle ~ \langle\! \langle \Delta_2| g(x_2) |\Delta_2, {\bf m}_2 \rangle ~ C^{\Delta_1, \Delta_2; \Delta_3}_{{\bf m}_1 ,{\bf m}_2; {\bf m}_3} \langle \Delta_3, {\bf m}_3 | g^{-1}(x_3) |\Delta_3 \rangle\!\rangle
\eea
We now show that this quantity satisfies the conformal Ward identity. To see this we note the following identities \cite{Bhatta:2016hpz}:
\bea
\label{gmmig}
g(x) \, M_{AB} &=& l^\mu_{AB} (x) \partial_\mu g(x) + \frac{1}{2} M_{bc} g(x) \left[ 
\omega_\mu^{bc} (x) l^\mu_{AB} (x) + {(R[g(x)])^{bc}}_{AB} \right] \cr
 M_{AB} g^{-1} (x) &=& -l^\mu_{AB} (x) \partial_\mu g^{-1} (x) + \frac{1}{2} \left[ 
\omega_\mu^{bc} (x) l^\mu_{AB} (x) + {(R[g(x)])^{bc}}_{AB} \right]  g^{-1} (x) M_{bc}
\eea
where the $l^\mu_{AB} (x)$ are the components of the Killing vector of the background geometry (\ref{eadsdplus1}) carrying the indices of the corresponding $so(1, d+1)$ algebra generator $M_{AB} \in \{M_{0a}, M_{ab}\}$ of the left hand side. Next we consider:
\bea
&& \langle\! \langle \Delta_1| g(x_1) M_{AB} |\Delta_1, {\bf m}_1 \rangle ~ \langle\! \langle \Delta_2| g(x_2) |\Delta_2, {\bf m}_2 \rangle \, C^{\Delta_1, \Delta_2; \Delta_3}_{{\bf m}_1 ,{\bf m}_2; {\bf m}_3} \langle \Delta_3, {\bf m}_3 | g^{-1}(x_3) |\Delta_3 \rangle\!\rangle \cr \cr
&& + \langle\! \langle \Delta_1| g(x_1) |\Delta_1, {\bf m}_1 \rangle ~ \langle\! \langle \Delta_2| g(x_2) M_{AB}  |\Delta_2, {\bf m}_2 \rangle \, C^{\Delta_1, \Delta_2; \Delta_3}_{{\bf m}_1 ,{\bf m}_2; {\bf m}_3} \langle \Delta_3, {\bf m}_3 | g^{-1}(x_3) |\Delta_3 \rangle\!\rangle \cr
&& \cr
&& - \langle\! \langle \Delta_1| g(x_1) |\Delta_1, {\bf m}_1 \rangle ~ \langle\! \langle \Delta_2| g(x_2)   |\Delta_2, {\bf m}_2 \rangle \, C^{\Delta_1, \Delta_2; \Delta_3}_{{\bf m}_1 ,{\bf m}_2; {\bf m}_3} \langle \Delta_3, {\bf m}_3 | M_{AB} g^{-1}(x_3) |\Delta_3 \rangle\!\rangle 
\eea
which vanishes identically as a consequence of the recursion relation (\ref{cgcrecursion}) for the CGC. On the other hand using the identities (\ref{gmmig}) above and the fact that the scalar cap is killed by $M_{ab}$'{\it s} we see that the OWN for the 3-point function (\ref{own3ptfn}) is invariant under simultaneous transformation of the three bulk points $(x_1, x_2, x_3)$ under any $AdS_{d+1}$ isometry. This in turn implies the conformal Ward identity in the limit of the external points $x_i$ approaching the boundary. It is of course true that the Ward identity completely determines the coordinate dependence of the 3-point function. Therefore, the question of finding the CGC is translated into finding expressions for the quantities $\langle\! \langle \Delta | g(x) |\Delta, {\bf m}\rangle$ and $\langle \Delta, {\bf m} | g^{-1}(x) |\Delta \rangle\!\rangle$ in the large radius limit, and then amputating them from the corresponding 3-point function (Fig. \!2).\footnote{Expressions of CGC for the scalar module obtained using this procedure can be found in appendix \ref{appA}.}
\vskip 1cm
\begin{center}
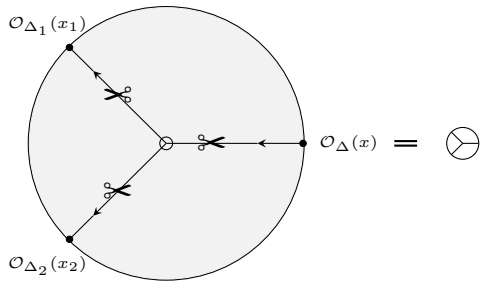

	\begin{tikzpicture} [scale=0.6]
	\draw [very thick] (-4,0) circle (3cm);
	\filldraw [gray!10] (-4,0) circle (3cm);
	\draw [>= stealth, ->] (-1,0) -- (-2,0);
	\draw (-2,0) -- (-4,0);
	\filldraw [black] (-1,0) circle (2pt);
	\node (scissors) at (-3,0) {\large \Rightscissors};
	\draw (-6.12,-2.12) -- (-5.6,-1.6);
	\draw [>= stealth, ->] (-4,0) -- (-5.6,-1.6);
	\node (scissors) at (-5.06,1.06) {\large \Leftscissors};
	\filldraw [black] (-6.12,-2.12) circle (2pt);
	\draw (-6.12,2.12) -- (-5.6,1.6);
	\draw [>= stealth, ->] (-4,0) -- (-5.6,1.6);
	\node (scissors) at (-5.06,-1.06) {\large \Rightscissors};
	\filldraw [black] (-6.12,2.12) circle (2pt);
	\draw (-4,0) circle (4pt);
	\draw [thick] (1,0.05) -- (1.5,0.05);
	\draw [thick] (1,-0.05) -- (1.5,-0.05);
	\draw (2.5,0) circle (10pt);
	\draw (2.5,0) -- (2.25,.25);
	\draw (2.5,0) -- (2.25,-.25);
	\draw (2.5,0) -- (2.85,0);
	%
	\node at (-6.6,-2.7) {\tiny $\mathcal{O}_{\Delta_2}(x_2)$};
	\node at (-6.6,2.6) {\tiny $\mathcal{O}_{\Delta_1}(x_1)$};
	\node at (0,0) {\tiny $\mathcal{O}_{\Delta}(x)$};
	\end{tikzpicture}
	\captionof{figure}{CG coefficients.}
\end{center}
\vskip .5cm
\subsection{Processing the Ingredients}
To proceed further we need the explicit expressions for the {\bf in-going} legs $\langle\! \langle \Delta | g(x) |\Delta, {\bf m}\rangle$ and the {\bf out-going}  legs $\langle \Delta, {\bf m} | g^{-1}(x) |\Delta \rangle\!\rangle$ which are matrix elements of $g(x)$ and $g^{-1}(x)$ between the cap states $|\Delta \rangle \! \rangle$ and normalised basis elements $|\Delta, {\bf m} \rangle$ of the scalar module. So we turn to finding a suitable orthonormal basis for the module over a scalar primary $|\Delta \rangle$ next. 
\vskip .5cm
\noindent\underline{\bf Scalar Module for $d\ge2$}
\vskip .5cm
The descendent states take the form $|\Delta, \{k_1, k_2, \dots ,k_d\} \rangle \sim \prod_{i=1}^{d}P_i^{k_i} |\Delta \rangle$. These states are eigenstates of the dilatation operator $D$ with eigenvalue $\Delta + \sum_{i=1}^d k_i$. States with different eigenvalues of $D$ are orthogonal. The set of states  with a given conformal weight form a reducible representation of the rotation algebra $so(d)$ -- which can be decomposed into a sum of irreps of $so(d)$. Then states belonging to different irreps will also be orthogonal. Therefore, a more suitable basis to work with would be in terms of the hyperspherical harmonics of the boundary $so(d)$ rotation algebra, $ ( P^2 )^s M^l_{{\bf m}} ({\bf P}) |\Delta\rangle $ where ${\bf m}$ denotes $(m_{d-2}, \cdots, m_2, m_1)$, whose conformal dimension is $\Delta + l + 2s$. In the rest of the paper we follow the conventions of \cite{WenAvery, Junker} for hyperspherical functions.\footnote{It turns out that this choice is responsible for giving the CPW{\it s} as a sum over contributions of given spin $l$, namely the Gegenbauer polynomial basis.} We define orthonormal states in this basis as follows\footnote{Note that the $so(d)$ symmetry dictates that the normalisation of these states do not depend on {\bf m}.}
\bea
( P^2 )^s M^l_{{\bf m}} ({\bf P}) |\Delta\rangle &:=& A_{l,s} \, | \Delta; \{ l,{\bf m},s\} \rangle \\
\langle \Delta | (K^2)^s \, {M^l_{{\bf m}}}^\star ({\bf K}) &:=& A^*_{l,s} \, \langle \Delta; \{ l,{\bf m},s\} | 
\eea
with 
\bea
\langle \Delta; \{ l',{\bf m}',s'\} | \Delta; \{ l,{\bf m},s\} \rangle = \delta_{ll'} \delta_{{\bf m} {\bf m}'} \delta_{ss'}
\eea
To find the normalisation $A_{l,s}$ let us start with the following observation 
\bea
\label{obsiden}
\langle \Delta|  e^{{\bf y} \cdot {\bf K}} e^{{\bf x} \cdot {\bf P}} |\Delta \rangle = \frac{1}{(1-2 \, {\bf x} \cdot {\bf y} + x^2 \, y^2)^\Delta}  ~ .
\eea
On the left hand side of the above identity we expand the plane waves $e^{{\bf x} \cdot {\bf P}}$ in terms of spherical waves:\footnote{Even though this formal expansion looks odd as it apparently depends not only on $P$ whose square is ${\bf P} \cdot {\bf P}$, but also appears in the denominator of the argument of the Gegenbauer polynomial --  we will shortly see that this is not a problem once interpreted correctly.}
\bea
\label{swe}
e^{{\bf x} \cdot {\bf P}} =  \sum_{l=0}^\infty  (2 l + d - 2)(d-4)!! \,j_{l}^{d}(x \, P) \, C_{l}^{\frac{d-2}{2}}\left(\frac{{\bf x} \cdot {\bf P}}{x \, P}\right)
\eea
where $j_l^{d}(x)$ is the spherical Bessel function and $C_l^{\mu} (z)$ is the Gegenbauer polynomials as defined below
\bea
C_{l}^{\frac{d-2}{2}}(x) =\frac{1}{(d-4)!!} \sum_{k=0}^{[l/2]} (-1)^k \frac{(2l-2k+d-4)!!}{(2k)!! (l-2k)!!} x^{l-2k}
\eea
and 
\bea
j_{l}^{d}(x) = \sum_{s=0}^\infty \frac{(-1)^s(x)^{l+2 s}}{(2 s)!! (d+ 2 l+2 s-2)!!} \, .
\eea
One can also write Gegenbauer polynomials in terms of hyperspherical harmonics using the well known identity
\bea
\label{gptosh}
\sum_{{\bf m}} Y _{l; {\bf m}}^{*}(\Omega_{x}) Y_{l;{\bf m}}(\Omega_{y})=\frac{\Gamma[\tfrac{d-2}{2}](2l+d-2)}{4 \pi^{d/2}} C_{l}^{\frac{d-2}{2}}\left(\frac{\vec x \cdot \vec y}{x \, y}\right)
\eea
Substituting these into the \eqref{swe} we get:
\bea
e^{{\bf x} \cdot {\bf P}}  =   4 \, a  \, \pi^{\frac{d}{2}} \sum_{l=0}^\infty \sum_{s=0}^\infty \frac{(x^{2})^{s}}{2^{l+2s} s! \Gamma[l+s+\frac{d}{2}]}\sum_{{{\bf m}}} M_{{\bf m}}^{l *}({\bf x})M_{{\bf m}}^{l}({\bf P})(P^{2})^{s}
\eea
where $M_{{\bf m}}^{l}({\bf x}) = x^l \, Y_{l;{\bf m}}(\Omega_{x}) $ and  
\bea
a =\begin{cases}
\frac{1}{2^{(d-2)/2}\Gamma[\frac{d-2}{2}]}, &\text{if $d$ is even} \\
\frac{\sqrt{\pi}}{2^{(d-1)/2}\Gamma[\frac{d-2}{2}]} , &\text{if $d$ is odd}
\end{cases}
\eea
Similarly
\bea
e^{{\bf y} \cdot {\bf K}} = 4\ a  \ \pi^{\frac{d}{2}} \sum_{l=0}^\infty \sum_{s=0}^\infty \frac{(y^{2})^{s}}{2^{l+2s} s! \Gamma[l+s+\frac{d}{2}]}\sum_{{{\bf m}}} M_{{\bf m}}^{l *}({\bf y})M_{{\bf m}}^{l}({\bf K})(K^{2})^{s} \,.
\eea
Therefore the left hand side of \eqref{obsiden} takes the following form
\bea
\label{lhs}
\langle \Delta|  e^{{\bf y} \cdot {\bf K}} e^{{\bf x} \cdot {\bf P}} |\Delta \rangle  =  \sum_{l=0}^\infty  \sum_{s=0}^\infty (x^2)^s (y^2)^{s}  \sum_{{\bf m}} M^{l}_{{\bf m}} ({\bf y}) \, M^{l *}_{{\bf m}} ({\bf x}) ~ 
|A_{l,s}|^2 \left(\frac{4\ a  \ \pi^{\frac{d}{2}}}{s!  \ 2^{l+2s} \ \Gamma(l+s+d/2)}\right)^2 \nonumber \\
\eea
Next we want to expand the {\it rhs} of (\ref{obsiden}) in the same basis. For this we first write 
\bea
\frac{1}{(1-2 \, {\bf x} \cdot {\bf y} + x^2 \, y^2)^\Delta} = \frac{1}{(1-2 \,  \xi \, t + t^2)^\Delta} 
\eea
with $t = x \, y$ and $\xi = t^{-1} {\bf x} \cdot {\bf y}$. We would now like to expand this quantity in terms of Gegenbauer polynomials $C_n^{\mu}(x)$. Luckily this exercise was done in~\cite{Cohl2013} which reads\footnote{This is a remarkable generalisation of how the Gegenbauer Polynomials $C_k^\mu(x)$ are defined through its generating function when $\Delta = \mu$.}
\bea
\label{cohlformula}
\frac{1}{(1-2 \, \xi \, t + t^2)^\Delta} = \frac{\Gamma(\mu)}{\Gamma(\Delta)} \sum_{k=0}^\infty C_k^\mu (\xi) \, t^k ~ \frac{\Gamma(\Delta + k)}{\Gamma(\mu+k)} ~~ {}_2F_1(\Delta+k, \Delta-\mu; \mu+k+1; t^2)
\eea
However, we are interested in expanding the {\it lhs} of (\ref{cohlformula}) in $d$-dimensional hyperspherical harmonics in ${\bf x}$ which requires us to choose $\mu = (d-2)/2$. Using  the series representation of the hypergeometric function:
\bea
{}_2F_1(\Delta+k, \Delta-\mu; \mu+k+1; t^2) = \frac{\Gamma(\mu+k+1)}{\Gamma(\Delta+k) \Gamma(\Delta - \mu)} \sum_{n=0}^\infty \frac{\Gamma(\Delta+k+n) \Gamma(\Delta - \mu +n)}{\Gamma(\mu+k +n+1)} ~ \frac{t^{2n}}{n!} \nonumber \\
\eea
and using the identity \eqref{gptosh} we finally arrive at
\bea
\label{rhs}
\frac{1}{(1-2 \, {\bf x \cdot y} + x^2 \, y^2)^\Delta} \!=\!  \frac{4 \  \pi^{\frac{d}{2}}}{\Gamma(\Delta)} \sum_{l,s=0}^\infty  \! \frac{\Gamma(\Delta+l+s) \Gamma(\Delta +s -\frac{d-2}{2})}{\Gamma(l +s+d/2) \, s!}  \, {(x^2)}^{l+2s} {(y^2)}^{l+2s} \! \sum_{{\bf m}}M^{l}_{{\bf m}} ({\bf y}) \, M^{l *}_{{\bf m}} ({\bf x}) \nonumber \\
\eea
Comparing \eqref{lhs} with \eqref{rhs}, we get\footnote{While this work was in progress \cite{Terashima:2017gmc} appeared where the same result was obtained in a different context.}
\bea
|A_{l,s} |^2 = \frac{2^{2l+4s}  \Gamma\left[l+s+d/2\right] \Gamma[\Delta+l+s]\Gamma \left[\Delta +s -\frac{(d-2)}{2} \right] s!}{4 \ a^2 \  \pi^{\frac{d}{2}}\Gamma[d/2] \Gamma[\Delta] \Gamma \left[\Delta-\frac{(d-2)}{2} \right]}
\eea
Having found an orthonormal basis for the scalar module we would like to now compute the legs (conformal wave functions) as described in the beginning of this section. 
\newpage
\vskip .5cm
\noindent\underline{\bf In-going legs: } ~~~ For this we start with $g(x) = e^{\rho D} e^{- {\bf x \cdot P}}$.
Then

\bea
&& \langle\!\langle \Delta |g(x)|\Delta ; \{l, {\bf m}, s\}\rangle \cr 
\cr
&& =\sum_{n=0}^{\infty} (-1)^n C_n \langle \Delta | \,(K^2)^n\, e^{\rho D} e^{- {\bf x \cdot P}} |\Delta ; \{l, {\bf m}, s\}\rangle \cr 
\cr
&&= \sum_{n=0}^{\infty} \frac{(-1)^n C_n}{A_{l,s}} A_{0,n}^*\,\langle \Delta ,\{0,0,n\}| \, e^{\rho D} e^{- {\bf x \cdot P}} \, ({P^2})^s \, M^l_{\bf m} ({\bf P}) |\Delta \rangle \cr 
\cr
&&=\frac{4 a \pi^{\frac{d}{2}}}{A_{l,s}} \sum_{n=0}^{\infty} (-1)^n C_n A_{0,n}^*\sum_{l'=0}^{\infty}  \sum_{s'=0}^{\infty} \frac{(x^2)^{s'}}{s'! \, 2^{l'+2s'}\,\Gamma(l'+s'+d/2)} \sum_{{\bf m}'} 
M^{l' *}_{{\bf m}'}(-{\bf x})  \cr
\cr 
&& \qquad \qquad \qquad \qquad \qquad \qquad \qquad \qquad  \times \,\langle \Delta ,\{0,0,n\}|\, e^{\rho D} (P^2)^{s+s'} M^{l'}_{{\bf m}'}({\bf P})M^{l}_{m}({\bf P}) |\Delta \rangle \nonumber 
\eea
 Now using the identity for the hyperspherical harmonics 
 \bea
 \label{eq:sp_harmonics_decomp}
 M^{l}_{\bf m}({\bf P}) M^{l'}_{\bf m'}({\bf P}) = \sum_{L} \sum_{\bf n} \,
 \begin{bmatrix}
 	l & l' & L \\
 	{\bf m} & {\bf m'} & {\bf n}
 \end{bmatrix}
  \, (P^2)^{\frac{l+l'-L}{2}} M^L_{\bf n}({\bf P}) 
 \eea
 where $ \big[ \begin{smallmatrix} l & l' & L \\ {\bf m} & {\bf m'} & {\bf n} \end{smallmatrix}\big] $ is $so(d)$ CG coefficients, we find 
 \begin{align}
 &\langle\!\langle \Delta |g(x)|\Delta ; \{l, {\bf m}, s\}\rangle \cr 
 &= \frac{4a\pi^{\frac{d}{2}}}{A_{l,s}} \sum_{n=0}^{\infty} (-1)^n C_n A^*_{0,n}\sum_{l'=0}^{\infty}  \sum_{s'=0}^{\infty} \frac{(x^2)^{s'}}{s'! \, 2^{l'+2s'}\,\Gamma(l'+s'+d/2)} \sum_{\bf m'} 
 M^{l' *}_{\bf m'}(-{\bf x})\,\,  e^{\rho(\Delta + l+l'+ 2(s+s'))} \cr
 &\qquad \qquad \qquad \qquad \qquad \qquad \times \sum_{L} \sum_{\bf n}
 \begin{bmatrix}
 l & l' & L \\
 {\bf m} & {\bf m'} & {\bf n}
 \end{bmatrix}
 \langle \Delta;\{0,0,n\}|(P^2)^{s+s'+ (l+l'-L)/2} M^L_{\bf n}({\bf P})|\Delta \rangle \cr 
 &= \frac{4a\pi^{\frac{d}{2}}}{A_{l,s}} \sum_{n=0}^{\infty} (-1)^n C_n A^*_{0,n}\sum_{l'=0}^{\infty}  \sum_{s'=0}^{\infty} \frac{(x^2)^{s'}}{s'! \, 2^{l'+2s'}\,\Gamma(l'+s'+d/2)} \sum_{\bf m'} 
 M^{l' *}_{\bf m'}(-{\bf x})\,\,  e^{\rho(\Delta + l+l'+ 2(s+s'))} \cr
 &\qquad \qquad \qquad \qquad \qquad \qquad \times \sum_{L} \sum_{\bf n}
 \begin{bmatrix}
 l & l' & L \\
 {\bf m} & {\bf m'} & {\bf n}
 \end{bmatrix}
 \, A_{L,s+s'+\frac{l+l'-L}{2}} \, \delta_{L0} \, \delta_{{\bf n}0} \, \delta_{n\left(s+s'+\frac{l+l'-L}{2} \right)}\nonumber 
 \end{align}
Carrying out the summation over $L$ and ${\bf n}$ we find
 \begin{align}
  &\langle\!\langle \Delta |g(x)|\Delta ; \{l, {\bf m}, s\}\rangle \cr 
  &= \frac{4a\pi^{\frac{d}{2}}}{A_{l,s}} \sum_{n=0}^{\infty} (-1)^n C_n A^*_{0,n}\sum_{l'=0}^{\infty}  \sum_{s'=0}^{\infty} \frac{(x^2)^{s'}}{s'! \, 2^{l'+2s'}\,\Gamma(l'+s'+d/2)} 
   M^{l}_{\bf m}(-{\bf x})\,\,  e^{\rho(\Delta + l+l'+ 2(s+s'))} \cr
  &\qquad \qquad \qquad \qquad \qquad \qquad \qquad \qquad \qquad \qquad \qquad \qquad \times \delta_{ll'} \, A_{0,s+s'+\frac{l+l'}{2}} \, \delta_{n\left(s+s'+\frac{l+l'}{2} \right)}
 \end{align}
 where we have used
 \bea
 \sum_{\bf m'} \,
 \begin{bmatrix}
 	l & l' & 0 \\
 	{\bf m} & {\bf m'} & 0
 \end{bmatrix} \, M^{l' *}_{\bf m'}({\bf x})
  = \delta_{ll'} \, M^{l}_{\bf m}({\bf x}) \, .
 \eea
 Therefore
 \bea
  && \langle\!\langle \Delta |g(x)|\Delta ; \{l, {\bf m}, s\}\rangle \cr 
  && \cr
  && = \frac{4a\pi^{\frac{d}{2}}}{A_{l,s}} \sum_{n=0}^{\infty} (-1)^n C_n A^*_{0,n}  \sum_{s'=0}^{\infty} \frac{(x^2)^{s'}}{s'! \, 2^{l+2s'}\,\Gamma(l+s'+d/2)} 
  M^{l }_{\bf m}(-{\bf x})\,\,  e^{\rho(\Delta + 2(l+s+s'))}  \, A_{0,s+s'+l} \, \delta_{n\left(s+s'+l \right)} \cr 
  && \cr
  &&= e^{\rho\Delta} \,\frac{4a\pi^{\frac{d}{2}}}{A_{l,s}} \, (x^2)^{-l-s} \, M^{l}_{\bf m}(-{\bf x}) \sum_{n=0}^{\infty} (-1)^n C_n |A_{0,n}|^2 \frac{\left(e^{2\rho} \, x^2\right)^{n}}{(n-s-l)! \, 2^{2n-2s-l}\,\Gamma(n-s+d/2)} \cr 
  && \cr
  &&=e^{-\rho\Delta} \frac{ 4 a \pi^{d/2} \, 2^{l+2s} }{A_{l,s} }\times  M^{l }_{\bf m}(-{\bf x})\times (-1)^{s+l} \,  (e^{2\rho})^{\Delta+l+s}(l+d/2)_s \, (\Delta)_{l+s} \cr \cr
  && \qquad \qquad \qquad \qquad \qquad \qquad \qquad \times \,  _2F_1 \left(\,\Delta+l+s,\, l+s+d/2;\, l+d/2;\, -e^{2\rho}x^2 \right)
 \eea
 Now we want to take $\rho \to \infty $ limit. We rewrite the hypergeometric function in the above expression using the identity
 \bea
 _2F_1(\, a,\, b;\, c;\, z) = (1-z)^{-a} \, _2F_1\left(\, a,\,c- b;\, c;\, \frac{z}{z-1}\right)
 \eea
 as
 \bea
 && _2F_1(\, \Delta+l+s,\, l+s+d/2;\, l+d/2;\, -e^{2\rho}x^2)  \cr 
 &&\qquad \qquad \qquad  = (1+e^{2\rho}x^2)^{-\Delta-l-s} \, _2F_1\left(\,\Delta+l+s,\, -s;\, l+\frac{d}{2};\, \frac{e^{2\rho}x^2}{1+e^{2\rho}x^2}\right)
 \eea
In the $\rho \to \infty $ limit the argument of the hypergeometric function tends to unity. As the following identity holds
 \bea
 _2F_1 (-n,\, b;\, c; 1 ) = \frac{(c-b)_n}{(c)_n} = \frac{\Gamma(c-b+n)\Gamma(c)}{\Gamma(c-b)\Gamma(c+n)} \, ,
 \eea
 to the leading order in $e^{\rho}$ the in-going leg becomes
 \bea
 \langle\!\langle \Delta |g(x)|\Delta ; \{l, {\bf m}, s\}\rangle \rightarrow
  e^{-\rho\Delta} \frac{ 4 a \pi^{d/2} \, 2^{l+2s} }{A_{l,s} }~ (-1)^{s+l} \, M^{l}_{\bf m}(-{\bf x})\, (x^2)^{-\Delta-l-s} \,(\Delta)_{l+s} \,(d/2-\Delta-s)_s + \cdots \nonumber
 \eea
where dots are subleading terms in $\rho \rightarrow \infty$ limit. Finally we use  
 %
$ (-x)_n = (-1)^n(x-n+1)_n$
 %
 and $ (-1)^l M^{l}_{\bf m} ({\bf x}) =  M^{l}_{\bf m} (-{\bf x}) $ to get
 \bea
 \label{finalincoming}
 \lim_{\rho \rightarrow \infty} e^{\rho\Delta}\langle\!\langle \Delta |g(x)|\Delta ; \{l, {\bf m}, s\}\rangle =
  4 \, a \, \pi^{d/2} \,  \frac{  2^{l+2s} }{A_{l,s} } ~  (\Delta)_{l+s} \,\left(\Delta-\frac{d-2}{2}\right)_s \, (x^2)^{-\Delta-l-s} \,M^{l}_{\bf m}({\bf x})\,  \nonumber \\
  \eea
 \newpage
\vskip .5cm
\noindent\underline{\bf Out-going legs:} ~~ For this we start with
%
$g^{-1}(y) = e^{{\bf y\cdot P}} e^{-\rho D}$,
%
and compute
\bea
&& \langle \Delta; \{l,{\bf m},s\}|g^{-1}(y)|\Delta \rangle\!\rangle \cr 
&&=  \sum_{n=0}^{\infty} (-1)^n C_n \,e^{-\rho(\Delta+2n)} \langle \Delta; \{l,{\bf m},s\}|e^{{\bf y\cdot P}}(P^2)^n|\Delta \rangle \cr 
&&= 4a\pi^{d/2} \sum_{n=0}^{\infty} (-1)^n C_n \,e^{-\rho(\Delta+2n)} \sum_{l'=0}^{\infty} \sum_{s'=0}^{\infty} \frac{(y^2)^{s'}}{s'! \, 2^{l'+2s'}\,\Gamma(l'+s'+d/2)} \sum_{\bf m'} 
M^{l' *}_{ \bf m'}({\bf y}) \cr 
&& \qquad \qquad \qquad \qquad \qquad \qquad \qquad \qquad \qquad \qquad \qquad \qquad \times A_{l',s'+n}  \, \delta_{ll'} \, \delta_{\bf mm'} \, \delta_{s(s'+n)} \cr 
&&=  4a\pi^{d/2} \sum_{n=0}^{\infty} (-1)^n C_n\, A_{l,s} \,e^{-\rho(\Delta+2n)} \, \frac{(y^2)^{s-n}}{(s-n)! \, 2^{l+2(s-n)}\,\Gamma(l+s-n+d/2)}  
M^{l *}_{\bf m}({\bf y}) \cr 
&&=e^{-\rho\Delta} \frac{4a\pi^{d/2}}{2^{l+2s}}\, A_{l,s} \sum_{n=0}^{\infty} (-1)^n C_n\, e^{-2n\rho} \frac{2^{2n}\,(y^2)^{s-n}}{(s-n)!\, \Gamma(l+s-n+d/2)}  \,M^{l *}_{\bf m}({\bf y})
\eea
As $\rho \to \infty $, to the leading order only the $n=0$ term contributes, so that we have the result
\bea
\label{finaloutgoing}
\lim_{\rho \rightarrow \infty}  e^{\rho \,\Delta} \, \langle \Delta; \{l,{\bf m},s\}|g^{-1}(y)|\Delta \rangle\!\rangle =\frac{4a\pi^{d/2}}{2^{l+2s}}\, A_{l,s} \, \frac{(y^2)^{s}}{(s)!\, \Gamma(l+s+d/2)}  \,M^{l *}_{\bf m}({\bf y})
\eea
The results of these rather lengthy, albeit straightforward exercises are (\ref{finalincoming}, \ref{finaloutgoing}). These two sets of functions (\ref{finalincoming}) and (\ref{finaloutgoing}) provide a representation and its conjugate representation respectively of the conformal algebra $so(1,d+1)$, on which the conformal generators $\{D, M_{\alpha\beta}, P_\alpha, K_\alpha\}$ act through their differential operator representations on scalar primaries with dimension $\Delta$. One can use these to derive matrix representations of the conformal generators and therefore, can be more appropriately called the {\it conformal wave functions}.

Finally let us quickly carry out a check on our conformal wave functions, namely, that when they are used in our OWN prescription they have to reproduce the appropriate two-point function for the scalar primaries. According to our prescription the two-point function can be obtained as
\bea
\langle \mathcal{O}_{\Delta} ({\bf x})  \mathcal{O}_{\Delta} ({\bf y}) \rangle 
&&= \lim_{\rho \to \infty}  e^{2\Delta\rho}\langle\!\langle \Delta | g(x) g^{-1}(y) |\Delta \rangle\!\rangle \cr 
&& = \lim_{\rho \to \infty}  e^{2\Delta\rho}\sum_{l=0}^{\infty} \sum_{s=0}^{\infty}\sum_{\bf m} \langle\!\langle \Delta |g(x)|\Delta ; \{l, {\bf m}, s\}\rangle \langle \Delta; \{l,{\bf m},s\}|g^{-1}(y)|\Delta \rangle\!\rangle
\eea
\begin{center}
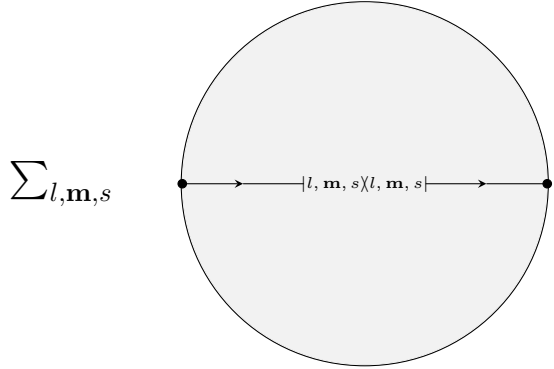

	\begin{tikzpicture}[scale=0.8]
	\draw [very thick] (0,0) circle (3cm);
	\filldraw [gray!10] (0,0) circle (3cm);
	\draw [>= stealth, ->] (-3,0) -- (-2,0);
	\draw (-2,0) -- (-1,0);
	\filldraw [black] (-3,0) circle (2pt);
	\draw [>= stealth, ->] (1,0) -- (2,0);
	\draw (2,0) -- (3,0);
	\filldraw [black] (3,0) circle (2pt);
	\node at (0.5,0) {\tiny $\langle l,{\bf m},s |$ };
	\node at (-0.5,0) {\tiny $ | l,{\bf m},s \rangle $ };
	\node at (-5,0) { \large $ \sum_{l, {\bf m}, s}$};
	\end{tikzpicture}
	\captionof{figure}{2-point function}
\end{center}
As $\rho \to \infty$ the above diagram evaluates to
\bea
 && \sum_{l=0}^{\infty} \sum_{s=0}^{\infty}\sum_{m=-l}^{l} \langle\!\langle \Delta |g(x)|\Delta ; \{l, m, s\}\rangle \langle \Delta; \{l,m,s\}|g^{-1}(y)|\Delta \rangle\!\rangle \cr 
 &&=  e^{-2\Delta\rho} (x^2)^{-\Delta}(4 a \pi^{d/2})^2 \sum_{l=0}^{\infty} \sum_{s=0}^{\infty} \frac{(\Delta)_{l+s}\, \left(\Delta-\frac{d-2}{2}\right)_s}{\Gamma(l +s+d/2) \, s!} \left(\frac{y}{x}\right)^{2s} \,  (x^2)^{-l}\sum_{\bf m} M^{l}_{\bf m}({\bf x}) M^{l *}_{\bf m}({\bf y}) ~~~~~~
\eea
Finally using \eqref{gptosh} and comparing with \eqref{cohlformula} we obtain
\bea
 \langle \mathcal{O}_{\Delta} ({\bf x})  \mathcal{O}_{\Delta} ({\bf y}) \rangle &=& 4 a^2 \pi^{d/2} (x^2)^{-\Delta} \left(1-2\, \frac{{\bf x \cdot y}}{x^2} + \frac{y^2}{x^2}\right)^{-\Delta} \cr 
 \cr
 &&=4 a^2 \pi^{d/2} |{\bf x-y}|^{-2\Delta}
\eea
This is the expected result for two-point function (up to an overall constant factor - which can be gotten rid of by multiplying the cap states by appropriate overall factors).

\subsection{Introducing OPE module}
Finally we need to amputate the legs (\ref{finalincoming}, \ref{finaloutgoing}) we have found in the previous subsection from the correlation function of three scalar primaries to find the CGC we need. The explicit expressions adapted to our method are given in appendix \ref{appA}. However, to compute, for example, the 4-point conformal partial waves we need CGC{\it s} that are already connected to two legs at a time -- which is obtained easily by starting with an appropriate 3-point function and amputating only one leg. This object depends on the boundary coordinates where two of the primaries are inserted, and carries labels of basis vectors of the module of the third primary. This is a close cousin of the so called OPE block \cite{Czech:2016xec, Boer:2016pqk}, which we call the {\it OPE module}. 

These OPE modules can be characterised by two types of identities. To spell them out let us label the representations of the conformal algebra $so(1, d+1)$ of interest by $(\Delta, {\bf l})$ where $\Delta$ is the conformal dimension and ${\bf l}$ represents all the independent Casimirs of the representation. States in such a representation $R$ can be labelled by $(\Delta, {\bf l}; {\bf m}, s)$ where ${\bf m}$ is again a collective index of magnetic quantum numbers. It turns out there are two types of these OPE modules which we denote by ${\cal B}^{(\Delta_1, {\bf l}_1; {\bf x}_1),  (\Delta_2, {\bf l}_2; {\bf x}_2)}_{(\Delta_3, {\bf l}_3; {\bf m}_3, s_3)}$ and ${\cal B}_{(\Delta_1, {\bf l}_1; {\bf x}_1),  (\Delta_2, {\bf l}_2; {\bf x}_2)}^{(\Delta_3, {\bf l}_3; {\bf m}_3, s_3)}$. Then these OPE modules are supposed to satisfy the Ward identities:
\bea
\left({\cal L}_{x_1}[M_{AB}] + {\cal L}_{x_2}[M_{AB}]\right)
{\cal B}^{(\Delta_1, {\bf l}_1; {\bf x}_1),  (\Delta_2, {\bf l}_2; {\bf x}_2)}_{(\Delta, {\bf l}; {\bf m}, s)} &=& {{\cal M}_{(\Delta, {\bf l}; {\bf m}, s)}}^{(\Delta, {\bf l}; {\bf m}', s')} [M_{AB}] {\cal B}^{(\Delta_1, {\bf l}_1; {\bf x}_1),  (\Delta_2, {\bf l}_2; {\bf x}_2)}_{(\Delta, {\bf l}; {\bf m}', s')} \cr \cr
\left({\cal L}_{x_1}[M_{AB}] + {\cal L}_{x_2}[M_{AB}]\right)
{\cal B}_{(\Delta_1, {\bf l}_1; {\bf x}_1),  (\Delta_2, {\bf l}_2; {\bf x}_2)}^{(\Delta, {\bf l}; {\bf m}, s)} &=& -   {\cal B}_{(\Delta_1, {\bf l}_1; {\bf x}_1),  (\Delta_2, {\bf l}_2; {\bf x}_2)}^{(\Delta, {\bf l}; {\bf m}', s')} \, {{\cal M}_{(\Delta, {\bf l}; {\bf m}', s')}}^{(\Delta, {\bf l}; {\bf m}, s)} [M_{AB}] \nonumber \\
\eea
where we denote the differential operator representation and the matrix representation of the conformal generator $M_{AB}$ by ${\cal L}[M_{AB}]$ and ${\cal M} [M_{AB}]$ respectively. From these identities it is very easy to see that both types of OPE modules satisfy the corresponding conformal Casimir equations. 

For the scalar blocks of interest here, the two types of OPE modules can be obtained by amputating either an in-going (\ref{finalincoming}), or an out-going leg (\ref{finaloutgoing}) from the appropriate 3-point functions: $\langle {\cal O}_{\Delta_1} (x_1) {\cal O}_{\Delta_2} (x_2) {\cal O}_{\Delta} (x) \rangle$, $\langle {\cal O}_{\Delta} (x)  {\cal O}_{\Delta_3} (x_3) {\cal O}_{\Delta_4} (x_4)\rangle$. See Fig. (4) for a pictorial representation of this procedure. 
\vskip 1cm
\begin{center}
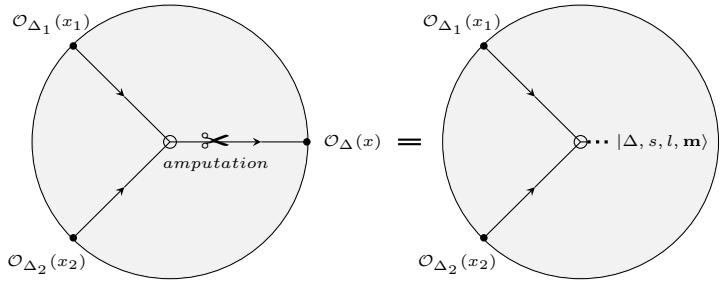

	\begin{tikzpicture}[scale =0.6]
	\draw [very thick] (-4,0) circle (3cm);
	\filldraw [gray!10] (-4,0) circle (3cm);
	\draw [>= stealth, ->] (-4,0) -- (-2,0);
	\draw (-2,0) -- (-1,0);
	\filldraw [black] (-1,0) circle (2pt);
	\node (scissors) at (-3,0) {\large \Rightscissors};
	\draw [>= stealth, ->] (-6.12,-2.12) -- (-5,-1);
	\draw (-5,-1) -- (-4,0);
	\filldraw [black] (-6.12,-2.12) circle (2pt);
	\draw [>= stealth, ->] (-6.12,2.12) -- (-5,1);
	\draw (-5,1) -- (-4,0);
	\filldraw [black] (-6.12,2.12) circle (2pt);
	\draw (-4,0) circle (4pt);
	\draw [thick] (1,0.05) -- (1.5,0.05);
	\draw [thick] (1,-0.05) -- (1.5,-0.05);
	\draw [very thick] (5,0) circle (3cm);
	\filldraw [gray!10] (5,0) circle (3cm);
	\draw [>= stealth, ->] (2.88,2.12) -- (4,1);
	\draw (4,1) -- (5,0);
	\draw [>= stealth, ->] (2.88,-2.12) -- (4,-1);
	\draw (4,-1) -- (5,0);
	\draw (5,0) circle (4pt);
	\draw (5,0) -- (5.15,0);
	\draw [very thick, dotted] (5.15,0) --  (5.65,0);
	\filldraw [black] (2.88,2.12) circle (2pt);
	\filldraw [black] (2.88,-2.12) circle (2pt);
	%
	\node at (-6.7,-2.7) {\tiny $\mathcal{O}_{\Delta_2}(x_2)$};
	\node at (-6.6,2.6) {\tiny $\mathcal{O}_{\Delta_1}(x_1)$};
	\node at (0,0) {\tiny $\mathcal{O}_{\Delta}(x)$};
	\node at (-3,-0.5) {\tiny $amputation$ };
	\node at (6.8,0) {\tiny $|\Delta, s, l, {\bf m}\rangle$};
	\node at (2.4,2.6) {\tiny $\mathcal{O}_{\Delta_1}(x_1)$};
	\node at (2.3,-2.7) {\tiny $\mathcal{O}_{\Delta_2}(x_2)$};
	\end{tikzpicture}
	\captionof{figure}{OPE module from 3-point function.}
\end{center}
Finally the method to obtain the 4-point conformal partial wave using the OWN prescription reduces to taking two types of OPE modules defined above and contracting the module indices.
\subsection{Computing the 4-point CPW}
Having equipped ourselves with all the ingredients needed, we now turn to compute four-point conformal blocks for scalar primaries of conformal weights $\Delta_i$ for $i = 1,2,3,4$. For simplicity we take the operator insertion points to be at ${\bf x_1} \to \infty,\, {\bf x_2} \to {\bf u} ,\, {\bf x_3} \to {\bf x}$ and $ {\bf x_4} \to {\bf 0}$ with ${\bf u} \cdot {\bf u} =1$. As elucidated above this four-point conformal block can be computed using two specific OPE modules.

One of the OPE modules we need can be extracted from the three-point function, with the operator insertions at $(\infty,\, {\bf u},\, {\bf y} ) $   by amputating the out-going leg anchored at the boundary-point ${\bf y}$. The corresponding OPE module is shown in the figure 5 below.
\vskip 1cm
\begin{center}
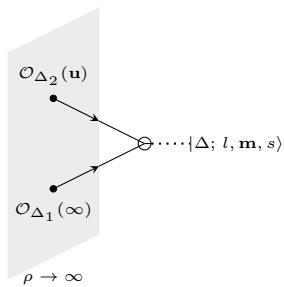

	\begin{tikzpicture} [scale=0.6]
	\filldraw [gray!15] (-1,2) -- (1,3) -- (1,-2) -- (-1,-3);
	\draw (2,0) -- (1,0.5);
	\draw [>= stealth, ->] (0,1) -- (1,0.5);
	\draw (2,0) -- (1,-0.5);
	\draw [>= stealth, ->] (0,-1) -- (1,-0.5);
	\filldraw [black] (0,1) circle (2pt);
	\filldraw [black] (0,-1) circle (2pt);
	\draw [thick, dotted] (2.15,0) -- (3,0);
	\draw (2,0) circle (4pt);
	\draw (2,0) -- (2.15,0);
	\node at (0,1.5) {\tiny $\small \mathcal{O}_{\Delta_2} ({\bf u})$};
	\node at (0,-1.5) {\tiny $\small \mathcal{O}_{\Delta_1} (\infty)$};
	\node at (0,-3) {\tiny $\rho \to \infty$ };
	\node at (4,0) {\tiny $|\Delta;\, l, { \bf m}, s \rangle$};
	\end{tikzpicture}
	\captionof{figure}{An OPE module}
\end{center}
The three-point function takes the form
\bea
 \langle \mathcal{O}_{\Delta_1}(\infty) \mathcal{O}_{\Delta_2} ({\bf u}) \mathcal{O}_{\Delta} ({\bf y}) \rangle &=& \lim_{z \to \infty}  (z^2)^{\Delta_1} \, \langle \mathcal{O}_{\Delta_1}({\bf z}) \mathcal{O}_{\Delta_2} ({\bf u}) \mathcal{O}_{\Delta} ({\bf y}) \rangle = \frac{1}{\left[ ({\bf u}- {\bf y})^2\right]^{\frac{\Delta_2+\Delta-\Delta_1}{2}}}
\eea
which can be expanded in terms of hyperspherical harmonics using \eqref{cohlformula} as
\bea
&& \langle \mathcal{O}_{\Delta_1}(\infty) \mathcal{O}_{\Delta_2} ({\bf u}) \mathcal{O}_{\Delta} ({\bf y}) \rangle \cr 
&& \cr 
&&= (4\pi^{d/2}) \sum_{l=0}^{\infty} \sum_{s=0}^{\infty} \frac{\left(\frac{\Delta_2+\Delta-\Delta_1}{2}\right)_{l+s} \, \left(\frac{\Delta_2+\Delta-\Delta_1}{2}- \frac{d-2}{2}\right)_{s}}{s! \, \Gamma(l+s+d/2)} \, (y^2)^s \sum_{\bf m} M^{l}_{{\bf m}} ({\bf u}) \, M^{l *}_{{\bf m}} ({\bf y})
\eea
Amputation of the out-going leg (\ref{finaloutgoing}) ending at ${\bf y}$ from the above expression gives 
\bea
\label{leftopemodule}
 \left[ \frac{4\pi^{d/2}}{s!\, (d/2)_{l+s} \, (\Delta)_{l+s} \, \left( \Delta-\frac{d-2}{2} \right)_s}\right]^{\frac{1}{2}} \,\left( \frac{\Delta-\Delta_{12}}{2}\right)_{l+s} \, \left( \frac{\Delta-\Delta_{12}}{2} - \frac{d-2}{2} \right)_s \, M^l_{{\bf m}} ({\bf u})
\eea
where $\Delta_{ij} \equiv \Delta_i - \Delta_j$. Similarly to find the other OPE module 
we start with the three-point function
\bea
 && \langle \mathcal{O}_{\Delta}({\bf y}) \mathcal{O}_{\Delta_3} ({\bf x}) \mathcal{O}_{\Delta_4} ({\bf 0}) \rangle 
 = (y^2)^{\frac{\Delta_3-\Delta_4-\Delta}{2}} (x^2)^{\frac{\Delta-\Delta_3-\Delta_4}{2}} \frac{1}{\left[ ({\bf y}- {\bf x})^2\right]^{\frac{\Delta+\Delta_3-\Delta_4}{2}}} 
\eea 
Expanding this in hyperspherical harmonics gives
\bea
 && \langle \mathcal{O}_{\Delta}({\bf y}) \mathcal{O}_{\Delta_3} ({\bf x}) \mathcal{O}_{\Delta_4} ({\bf 0}) \rangle \cr
 && \cr 
 &&= (4\pi^{d/2}) (x^2)^{\frac{\Delta-\Delta_3-\Delta_4}{2}} \sum_{l, s=0}^{\infty}  \frac{\left(\frac{\Delta+\Delta_3-\Delta_4}{2}\right)_{l+s} \, \left(\frac{\Delta+\Delta_3-\Delta_4}{2}- \frac{d-2}{2}\right)_{s}}{s! \, \Gamma(l+s+d/2)} \, (x^2)^s (y^2)^{\Delta-l-s} \sum_{\bf m} M^{l}_{{\bf m}} ({\bf y}) \, M^{l *}_{{\bf m}} ({\bf x}) \cr 
 &&
\eea
\begin{center}
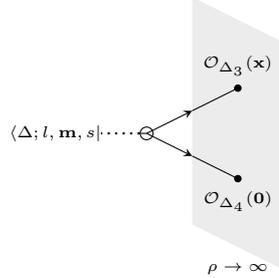

	\begin{tikzpicture} [scale=0.6]
	\filldraw [gray!15] (-1,3) -- (1,2) -- (1,-3) -- (-1,-2);
	\draw [>= stealth, ->] (-2,0) -- (-1,0.5);
	\draw (-1,0.5) -- (0,1);
	\filldraw [black] (0,1) circle (2pt); 
	\draw [>= stealth, ->] (-2,0) -- (-1,-0.5);
	\draw (-1,-0.5) -- (0,-1);
	\filldraw [black] (0,-1) circle (2pt);
	\draw (-2,0) circle (4pt);
	\draw (-2,0) -- (-2.15,0);
	\draw [thick, dotted] (-2.15,0) -- (-3,0);
	\node at (-4,0) {\tiny $\langle \Delta; l, {\bf m}, s |$};
	\node at (0,1.5) {\tiny $\small \mathcal{O}_{\Delta_3} ({\bf x})$};
	\node at (0,-1.5) {\tiny $\small \mathcal{O}_{\Delta_4} ({\bf 0})$};
	\node at (0,-3) {\tiny $\rho \to \infty$ };
	\end{tikzpicture}
	\captionof{figure}{Another OPE module.}
\end{center}
Now amputating the in-going leg (\ref{finalincoming}) starting from ${\bf y}$, we obtain 
\bea
\label{rightopemodule}
&& (x^2)^{\tfrac{(\Delta - \Delta_3-\Delta_4)}{2}} 
\left[ \tfrac{1}{ 4\pi^{d/2} \, s!\, (d/2)_{l+s} \, (\Delta)_{l+s} \, \left( \Delta-\frac{d-2}{2} \right)_s}\right]^{\frac{1}{2}} \,\tfrac{1}{\Gamma(d/2)}\left( \tfrac{\Delta+\Delta_{34}}{2}\right)_{l+s} \, \left( \tfrac{\Delta+\Delta_{34}}{2} - \tfrac{d-2}{2} \right)_s \,(x^2)^s M^{l *}_{{\bf m}} ({\bf x})\cr 
&&
\eea
Finally we glue the OPE modules (\ref{leftopemodule}) and (\ref{rightopemodule}) to compute the four-point conformal partial wave (see figure 7).
\vskip 1cm
\begin{center}
	\begin{tikzpicture} [scale=0.8]
	\draw [very thick] (-4,0) circle (3cm);
	\filldraw [gray!10] (-4,0) circle (3cm);
	\draw [very thick](4,0) circle (3cm);
	\filldraw [gray!10] (4,0) circle (3cm);
	\draw [thick] (-0.25,0.05) -- (0.25,0.05);
	\draw [thick] (-0.25,-0.05) -- (0.25,-0.05);
	\draw [>= stealth, ->] (-6.25,2) -- (-5.875,1);
	\draw (-5.875,1) -- (-5.5,0);
	\draw [>= stealth, ->] (-6.25,-2) -- (-5.875,-1);
	\draw (-5.875,-1) -- (-5.5,0);
	\draw (-5.5,0) circle (4pt);
	\draw (-5.5,0) -- (-5.45,0);
	\draw [very thick, dotted] (-5.45,0) -- (-5,0);
	\filldraw [black] (-6.25,2) circle (2pt);
	\filldraw [black] (-6.25,-2) circle (2pt);
	\draw [>= stealth, ->] (-2.5,0) -- (-2.125,1);
	\draw (-1.75,2) -- (-2.125,1);
	\draw [>= stealth, ->] (-2.5,0) -- (-2.125,-1);
	\draw (-1.75,-2) -- (-2.125,-1);
	\draw (-2.5,0) circle (4pt);
	\draw (-2.5,0) -- (-2.55,0);
	\draw [very thick, dotted] (-2.55,0) -- (-3,0);
	\filldraw [black] (-1.75,2) circle (2pt);
	\filldraw [black] (-1.75,-2) circle (2pt);
	\draw [>= stealth, ->] (5.5,0) -- (5.875,1);
	\draw (6.25,2) -- (5.875,1);
	\draw [>= stealth, ->] (5.5,0) -- (5.875,-1);
	\draw (6.25,-2) -- (5.875,-1);
	\draw (5.5,0) circle (4pt);
	\draw [>= stealth, ->] (2.5,0) -- (4,0);
	\draw  (4,0) -- (5.5,0);
	\filldraw [black] (6.25,2) circle (2pt);
	\filldraw [black] (6.25,-2) circle (2pt);
	\draw [>= stealth, ->] (1.75,2) -- (2.125,1);
	\draw (2.125,1) -- (2.5,0);
	\draw [>= stealth, ->] (1.75,-2) -- (2.125,-1);
	\draw (2.125,-1) -- (2.5,0);
	\draw (2.5,0) circle (4pt);
	\filldraw [black] (1.75,2) circle (2pt);
	\filldraw [black] (1.75,-2) circle (2pt);
	\node at (-3.5,0) {\tiny $\langle l, {\bf m},s  |$ };
	\node at (-4.5,0) {\tiny $| l, {\bf m},s  \rangle$ };
	\node at (-9,0) { \large $ \sum_{l, {\bf m},s }$};
	\node at (-6.6,2.6) {\tiny $\mathcal{O}_{\Delta_1} (x_1)$};
	\node at (-6.6,-2.6) {\tiny $\mathcal{O}_{\Delta_2} (x_2)$};
	\node at (-1.4,2.6) {\tiny $\mathcal{O}_{\Delta_4} (x_4)$};
	\node at (-1.4,-2.7) {\tiny $\mathcal{O}_{\Delta_3} (x_3)$};
	\node at (1.4,2.6) {\tiny $\mathcal{O}_{\Delta_1} (x_1)$};
	\node at (1.4,-2.7) {\tiny $\mathcal{O}_{\Delta_2} (x_2)$};
	\node at (6.6,-2.7) {\tiny $\mathcal{O}_{\Delta_3} (x_3)$};
	\node at (6.6,2.6) {\tiny $\mathcal{O}_{\Delta_4} (x_4)$};
	\node at (4,0.3) {\tiny $\mathcal{O}_{\Delta}$};
	\end{tikzpicture}
	\vskip .5cm
	
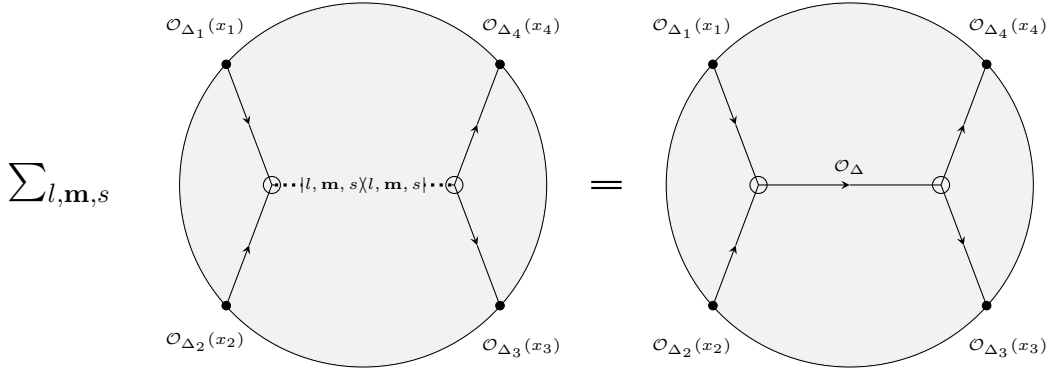
\captionof{figure}{4-point block from OPE modules.}
\end{center}
Thus our prescription for the corresponding four-point conformal partial wave gives
\bea
 W^{(d)}_{\Delta, 0} (\Delta_i, {\bf x}) &=&
 (x^2)^{\tfrac{(\Delta - \Delta_3-\Delta_4)}{2}} \frac{1}{\Gamma(d/2)} 
 \sum_{l,s}  \frac{\left( \tfrac{\Delta-\Delta_{12}}{2}\right)_{l+s} \, \left(\tfrac{\Delta+\Delta_{34}}{2}\right)_{l+s}}{s!\, (d/2)_{l+s} \, (\Delta)_{l+s} \, \left( \Delta-\frac{d-2}{2} \right)_s}  \cr
&\times& \left( \frac{\Delta-\Delta_{12}}{2} - \frac{d-2}{2} \right)_s \, \left( \frac{\Delta+\Delta_{34}}{2} - \frac{d-2}{2} \right)_s  (x^2)^s \, \, \sum_{{\bf m}} {M^l_{{\bf m}}}^\star ({\bf x}) \, M^l_{{\bf m}} ({\bf u}) \nonumber \\
\eea
Using \eqref{gptosh} we express this in terms of Gegenbauer polynomials
\bea
\label{ourblock}
W^{(d)}_{\Delta, 0} (\Delta_i, {\bf x}) &=&
 (x^2)^{\tfrac{(\Delta - \Delta_3-\Delta_4)}{2}} \frac{\Gamma\left(\tfrac{d-2}{2}\right) }{4\pi^{d/2} \, \Gamma(d/2)}
  \sum_{l,s}  \frac{(2l+d-2) \,\left( \tfrac{\Delta-\Delta_{12}}{2}\right)_{l+s} \, \left(\tfrac{\Delta+\Delta_{34}}{2}\right)_{l+s}}{s!\, (d/2)_{l+s} \, (\Delta)_{l+s} \, \left( \Delta-\frac{d-2}{2} \right)_s}  \cr
  &&~~~~\times \left( \frac{\Delta-\Delta_{12}}{2} - \frac{d-2}{2} \right)_s \, \left( \frac{\Delta+\Delta_{34}}{2} - \frac{d-2}{2} \right)_s  x^{l+2s} \,C_{l}^{\frac{d-2}{2}}\left(\frac{{\bf x \cdot u}}{x}\right) \nonumber \\
\eea
This is our final result for the scalar conformal partial wave.  Even though we assumed $d \ge 2$, we will see in the next section this result also holds for $d=1$. Notice that as advertised in the introduction our answer is naturally given in terms of Gegenbauer polynomials. 

A result for the same quantity already exists in the literature in terms of the cross ratios \cite{Dolan:2011dv}. In appendix \ref{appB} we show our answer agrees with their result.

In principle one can put together the conformal wave functions of section 2, and the CGC of appendix \ref{appA} suitably to generate the scalar CPW{\it s} of any higher-point scalar correlators as well (as was done for $d=2$ case in \cite{Bhatta:2016hpz}).
\section{Recovery of Results in $d \le 4$}
\label{sec3}
In this section we want to recover the known results for four-point scalar conformal partial waves in $d = 1, \cdots, 4$ from our answer (\ref{ourblock}). For this we find it  convenient to express our answer in different variables. Writing ${\bf x \cdot u} = x \, \cos \theta$, we  define 
\bea
\label{zbzdefs}
 z = x \, e^{i\theta},  \qquad \bar z = x \, e^{-i\theta} \, .
\eea
In terms of these variables $(z, \bar z)$  the four-point CPW (\ref{ourblock}) takes the form
\bea
\label{ourblockzbz}
 W^{(d)}_{\Delta, 0} (\Delta_i; z, \bar z) &&=
 (z\bar z)^{\tfrac{(\Delta - \Delta_3-\Delta_4)}{2}} \frac{1}{4\pi^{d/2}}
 \sum_{l,s}  \frac{(2l+d-2) \,\left( \tfrac{\Delta-\Delta_{12}}{2}\right)_{l+s} \, \left(\tfrac{\Delta+\Delta_{34}}{2}\right)_{l+s}}{s!\, (d/2)_{l+s} \, (\Delta)_{l+s} \, \left( \Delta-\frac{d-2}{2} \right)_s}  \cr \cr
 &&\!\!\!\!\!\times \left( \tfrac{\Delta-\Delta_{12}}{2} - \frac{d-2}{2} \right)_s \, \left( \frac{\Delta+\Delta_{34}}{2} - \frac{d-2}{2} \right)_s  (z\bar z)^{s+\frac{l}{2}} \,\frac{2}{(d-2)}C_{l}^{\frac{d-2}{2}}\left(\tfrac{z+\bar z}{2 \sqrt{z\bar z}}\right) 
\eea
\vskip .5cm
\noindent\underline{\bf $d=4$:} ~~~ Substituting $d=4$ in (\ref{ourblockzbz}) and manipulating further we find
\bea
 W^{(4)}_{\Delta, 0} (\Delta_i, z, \bar z) &=& \frac{1}{z-\bar z} \,(z \bar z)^{\frac{1}{2}(\Delta - \Delta_3-\Delta_4)}  \sum_{l,s=0}^\infty \Gamma(\tfrac{1}{2}(\Delta+ \Delta_{34})+l+s) \Gamma(\tfrac{1}{2}(\Delta- \Delta_{12})+l+s)  \cr
 && ~~~~~\times \Gamma(\tfrac{1}{2}(\Delta- \Delta_{12})+s-1)\Gamma(\tfrac{1}{2}(\Delta + \Delta_{34})+s-1) \cr
 && ~~~\times  \frac{(l+1) \Gamma(\Delta) \, \Gamma(\Delta-1)}{s! \, (l+s+1)! \Gamma(\Delta + l +s) \Gamma(\Delta+s-1)} \,   (z^{l+s+1} {\bar z}^s - z^s {\bar z }^{l+s+1}) \cr  \cr
 &=&  \,   \Gamma(\alpha) \, \Gamma(\alpha-1) \Gamma(\beta) \, \Gamma(\beta-1) \, \frac{1}{z-\bar z}  \, (z \bar z)^{\frac{1}{2}(\Delta - \Delta_3-\Delta_4)}    \cr
 && \!\!\!\!\!\!\!\!\!\!\!\!\!\!\!\!\!\!\!\!\!\!\!\!\!\!\!\! \left[ z ~ {}_2F_1(\alpha,\beta,\Delta, z) \, {}_2F_1(\alpha -1,\beta-1,\Delta-2, \bar z) - \bar z ~ {}_2F_1(\alpha,\beta,\Delta, \bar z) \, {}_2F_1(\alpha -1,\beta-1,\Delta-2, z) \right] \cr 
 &&
\eea
where $\alpha = \tfrac{1}{2}(\Delta- \Delta_{12})$ and $\beta = \tfrac{1}{2}(\Delta+ \Delta_{34})$. The details of the calculation of how to go from the first to the second expression are relegated to appendix \ref{appC}. Our answer perfectly matches with the known results \cite{Dolan:2011dv}.
\vskip .5cm
\noindent\underline{$d=3$:} ~~~
When $d=3$ the Gegenbauer polynomials used to express the answer (\ref{ourblockzbz}) become the Legendre polynomials, {\it i.e.,} $C^{1/2}_l (\cos\theta) = P_l(\cos\theta)$. Therefore, our answer reads
\bea
W^{(3)}_{\Delta, 0}(\Delta_i; z, \bar z) &&=
(z\bar z)^{\tfrac{(\Delta - \Delta_3-\Delta_4)}{2}} \frac{1}{\pi^{3/2}}
\sum_{l,s=0}^\infty  \frac{(l+1/2) \,\left( \tfrac{\Delta-\Delta_{12}}{2}\right)_{l+s} \, \left(\tfrac{\Delta+\Delta_{34}}{2}\right)_{l+s}}{s!\, (3/2)_{l+s} \, (\Delta)_{l+s} \, \left( \Delta-1/2 \right)_s}  \cr \cr
&&\times \left( \tfrac{\Delta-\Delta_{12}}{2} - \frac{1}{2} \right)_s \, \left( \frac{\Delta+\Delta_{34}}{2} - \frac{1}{2} \right)_s  (z\bar z)^{s+\frac{l}{2}} \,P_l\left(\tfrac{z+\bar z}{2 \sqrt{z\bar z}}\right) 
\eea
We are not aware of any closed form for this case. There exists a conjectured formula by \cite{Hogervorst:2016hal} where the $d=3$ four-point block is written in terms of 2$d$ blocks. We have checked that our answer also agrees with \cite{Hogervorst:2016hal} for large ranges of $l$ and $s$.
\vskip .5cm
\noindent\underline{$d=2$:} ~~~~ 
To recover the answer for $d=2$ we have to take the $d\rightarrow 2$ limit of (\ref{ourblockzbz}). We find
\bea
  W^{(2)}_{\Delta, 0}(\Delta_i; z, \bar z) 
 &=&
 (z\bar z)^{\tfrac{(\Delta - \Delta_3-\Delta_4)}{2}} \frac{1}{\pi}
 \sum_{l,s=0}^\infty  \frac{\left( \tfrac{\Delta-\Delta_{12}}{2}\right)_{l+s} \, \left(\tfrac{\Delta+\Delta_{34}}{2}\right)_{l+s}}{s!\, (l+s)! \, (\Delta)_{l+s} \, \left( \Delta \right)_s}  \, \left( \tfrac{\Delta-\Delta_{12}}{2}  \right)_s \, \left( \tfrac{\Delta+\Delta_{34}}{2} \right)_s  (z\bar z)^{s+\frac{l}{2}} \,\cos \, (l\theta) \cr 
  &=&  \frac{1}{2\pi} (z\bar z)^{\tfrac{(\Delta - \Delta_3-\Delta_4)}{2}} \sum_{l,s=0}^\infty \frac{\left( \tfrac{\Delta-\Delta_{12}}{2}\right)_{l+s} \, \left(\tfrac{\Delta+\Delta_{34}}{2}\right)_{l+s}}{ (l+s)! \, (\Delta)_{l+s}} \frac{\left( \tfrac{\Delta-\Delta_{12}}{2}  \right)_s \, \left( \tfrac{\Delta+\Delta_{34}}{2} \right)_s}{s! \, (\Delta)_s} \, (z^{l+s} \bar z^s + z^s \bar z^{l+s}) \nonumber
\eea
where we have used the following identity
\bea
 \lim_{\mu \to 0} \, \frac{1}{\mu} \, C^{\mu}_l(\cos \theta) = \frac{2}{l} T_l (\cos \theta) = \frac{2}{l} \cos(l\theta)
\eea
where $T_l$ are Chebyshev polynomials of the first kind. Finally performing the summations we recover the familiar answer for scalar CPW in two dimensions
\bea
W^{(2)}_{\Delta, 0} (\Delta_i; z, \bar z) &=& \frac{1}{\pi} (z\bar z)^{\tfrac{1}{2}(\Delta - \Delta_3-\Delta_4)}  {}_2F_1 \left[\left( \tfrac{\Delta-\Delta_{12}}{2}\right),\left(\tfrac{\Delta+\Delta_{34}}{2}\right); \, \Delta;\, z\right]{}_2F_1 \left[\left( \tfrac{\Delta-\Delta_{12}}{2}\right),\left(\tfrac{\Delta+\Delta_{34}}{2}\right); \, \Delta;\, \bar z\right] \cr 
&& ~~~~~~~~~ \qquad\qquad \qquad\qquad \qquad\qquad \qquad\qquad \qquad\qquad \qquad\qquad 
\eea
 \vskip .3cm
\noindent\underline{\bf $d=1$:} ~~~
This case corresponds to $\frac{d-2}{2} = -1/2$ and the corresponding Gegenbauer polynomials take the following form:
\bea
C^{-\frac{1}{2}}_l (\chi) = \delta_{l,0} - \chi \, \delta_{l,1} + \theta(l-2) \frac{1-\chi^2}{l(l-1)} \frac{d}{d\chi} P_{l-1}(\chi)
\eea
Further, in this case all the positions of the operators are simply real numbers. In particular, the unit vector ${\bf u}$ becomes either $1$ or $-1$. Without loss of generality we take ${\bf u}=1$. Then the argument of the Gegenbauer polynomials in \eqref{ourblock}, $ \hat{ {\bf x} }\cdot {\bf u}$ also becomes $\pm 1$ depending on the sign of ${\bf x}$. For both the cases the Gegenbauer polynomial simplifies to
\bea
 C^{(-1/2)}_l (\pm 1) =  \delta_{l0} \mp \delta_{l1}  = \delta_{l0} - {\rm sign} ({\bf x}) \, \delta_{l1}
\eea
Then the expression for 4-point CPW splits into two parts as follows
\bea
\label{onedblockone}
W^{(1)}_{\Delta, 0}(\Delta_i, x) &=&
(x^2)^{\tfrac{(\Delta - \Delta_3-\Delta_4)}{2}} \frac{1 }{2\sqrt{\pi}} \left[
\sum_{s=0}^{\infty}  \frac{\left( \alpha \right)_{s} \, \left(\beta\right)_{s}}{s!\, (1/2)_{s} \, (\Delta)_{s} \, \left( \Delta+\frac{1}{2} \right)_s}  \,\left( \alpha + \frac{1}{2} \right)_s \, \left( \beta + \frac{1}{2} \right)_s  x^{2s} \right. \cr 
 &&
 + ~ {\rm sign} (x) \,  \left.
 \sum_{s=0}^{\infty}  \frac{ \left( \alpha \right)_{s+1} \, \left(\beta\right)_{s+1}}{s!\, (1/2)_{s+1} \, (\Delta)_{s+1} \, \left( \Delta+\frac{1}{2} \right)_s}  \,\left( \alpha + \frac{1}{2} \right)_s \, \left( \beta + \frac{1}{2} \right)_s  x^{2s+1} \right]  
\eea
where $\alpha = \tfrac{1}{2}(\Delta- \Delta_{12})$ and $\beta = \tfrac{1}{2}(\Delta+ \Delta_{34})$ as before. Now using the following identities for Pochhammer symbols
\bea
 &&\left(A \right)_s \left( A + \frac{1}{2} \right)_s = \frac{1}{2^{2s}} \, \left( 2A\right)_{2s} \, , \qquad \left(A\right)_{s+1} \left(A + \frac{1}{2} \right)_s = \frac{1}{2^{2s+1}} \, \left( 2A\right)_{2s+1} \cr 
 &&
\eea
for $A \in \{\alpha, \beta, \Delta \}$, and
\bea
 s! \left(\frac{1}{2}\right)_{s} = \frac{(2s)!}{2^{2s}}, \qquad s! \left(\frac{1}{2}\right)_{s+1} = \frac{(2s+1)!}{2^{2s+1}}
\eea
we can show that the expression (\ref{onedblockone}) can be written as a single sum, which can be carried out to yield the answer
\bea
\label{1d4ptblock}
 W^{(1)}_{\Delta, 0}(x) &=& \frac{1}{2\sqrt{\pi}} ~ x^{\Delta - \Delta_3-\Delta_4} \, {}_2F_1 \left( 2\alpha , 2\beta; \, 2\Delta;\, x \right) 
\eea
where $x = |{\bf x}|$. This expression agrees with the known result \cite{Qiao:2017xif, Gross:2017aos} for the $d=1$ case.
\section{Seed Blocks and Recursion Relations}
\label{sec4}
There exist in the literature some powerful recursion relations that enable one to compute the CPW in a given dimension in terms of those in lower dimensions \cite{SimmonsDuffin:2012uy, Penedones:2015aga, Hogervorst:2016hal}. For instance, one such recursion relation among the even dimensional CPW{\it s} was given in \cite{SimmonsDuffin:2012uy}. In this section we give a different (and simpler) proof of this relation using our answer, and provide a counterpart of such a relation among the odd dimensional CPW{\it s}. For this we begin by extracting the conformal block from the CPW via the relation: $W^{(d)}_{\Delta, 0} (\Delta_i, {\bf x}):= (x^2)^{-\frac{1}{2}( \Delta_3+\Delta_4)} G^{\mu}_{\Delta}(\alpha,\beta; {\bf x})$ where $\alpha = \tfrac{1}{2}(\Delta- \Delta_{12}),\,
\beta  =\tfrac{1}{2}(\Delta+ \Delta_{34})$ and $\mu = \frac{d-2}{2}$. Then from (\ref{ourblock}) we have:
\bea
\label{defG}
G^{\mu}_{\Delta}(\alpha,\beta; {\bf x}) =   
&& (x^2)^{\frac{\Delta}{2}} \sum_{l,s=0}^\infty \frac{(\alpha)_{l+s} (\beta)_{l+s} (\alpha-\mu)_{s} (\beta-\mu)_{s}}{s!(\Delta)_{l+s}(\mu +1)_{l+s}} \left(1+\tfrac{l}{\mu}\right) x^{l+2s} \, \,C_{l}^{\mu}\left(\cos  \,\theta\right) \, .
\eea
 Differentiating with respect to $\cos\theta$ and using the identity
\bea
\frac{d}{dz} C_{l}^{\mu}\left(z\right) = 2 \mu \, C_{l-1}^{\mu+1}\left(z\right)
\eea
(\ref{defG}) becomes 
\bea
\frac{d G^{\mu}_{\Delta}(\alpha,\beta; {\bf x})}{d \, (\cos\theta)} &=&  
(x^2)^{\frac{\Delta}{2}} \sum_{l=1}^{\infty}\sum_{s=0}^{\infty} \frac{(\alpha)_{l+s} (\beta)_{l+s} (\alpha-\mu)_{s} (\beta-\mu)_{s}}{s!(\Delta)_{l+s}(\mu +1)_{l+s}} \left(1+\tfrac{l}{\mu}\right) x^{l+2s} \,2 \mu \,C_{l-1}^{\mu+1}\left(\cos\theta\right)\nonumber \cr
 &=& \frac{2 \alpha \beta}{\Delta}  (x^2)^{\frac{\Delta+1}{2}}
\sum_{l, s=0}^\infty \frac{(\alpha+1)_{l+s} (\beta+1)_{l+s} (\alpha-\mu)_{s} (\beta-\mu)_{s}}{s!(\Delta+1)_{l+s}(\mu +2)_{l+s}} \left(1+\tfrac{l}{\mu+1}\right) x^{l+2s} \,C_{l}^{\mu+1}\left(\cos\theta\right)\nonumber \\
&=& \frac{2 \alpha \beta}{\Delta} G^{\mu+1}_{\Delta+1}(\alpha+1,\beta+1; {\bf x}) 
\eea
where, in going from the first to the second line we have replaced $l \rightarrow l+1$  and used the identity: $(\alpha)_{n+1}=\alpha \, (\alpha+1)_n$. By applying this relation repeatedly (say $k$ times) we arrive at:
\bea
\label{recident}
G^{\mu}_{\Delta}(\alpha,\beta; {\bf x}) &=&  \frac{  (\Delta-k)_k}{2^k \, (\alpha-k)_k(\beta-k)_k} \left(\frac{d}{d\, \cos\theta}\right)^k G^{\mu -k}_{\Delta-k} (\alpha -k ,\beta -k; {\bf x}) \cr
&=&  \frac{(\alpha)_{-k} (\beta)_{-k}}{2^k \, (\Delta)_{-k}} \left(\frac{d}{d\, \cos\theta}\right)^k G^{\mu -k}_{\Delta-k} (\alpha -k ,\beta -k; {\bf x})
\eea
where we have used $(\alpha-k)_k \, (\alpha)_{-k} =1$. Since $\mu \rightarrow \mu+1$ corresponds to $d \rightarrow d+2$ the equation (\ref{recident}) says that we can get all even (odd) dimensional conformal blocks starting from, say the 2$d$ (3$d$) blocks. Writing $d=2k+2 + 2 \gamma$ where $\gamma = 0$ for even $d$ and $\gamma = 1/2$ for odd $d$, we can recast this result as
\bea
G^{k+\gamma}_{\Delta}(\alpha,\beta; {\bf x})=\frac{(\alpha)_{-k}(\beta)_{-k}}{(\Delta)_{-k}}\left(\frac{d}{d v}\right)^{k} G^\gamma_{\Delta - k} (\alpha - k ,\beta - k; {\bf x})
\eea
where using (\ref{zbzdefs}) we have defined
\bea
\frac{d}{d v}=\frac{x}{z-\bar{z}}\left(z\frac{d}{d z}-\bar{z}\frac{d}{d \bar{z}}\right) \, .
\eea
This result for the case of $\gamma = 0$ (relating different even dimensional blocks) is the one found in \cite{SimmonsDuffin:2012uy} -- whereas the case of $\gamma = 1/2$ is its odd dimensional counterpart.
\section{Some Odds and Ends}
\label{sec5}
In this section we present a couple of additional results that are a selection of possible generalisations in various directions of the cases considered so far. One of the limitations is the restriction to scalar operators (both in the external and the internal legs). The cases of $d=1$ and $d=2$ are the simplest to address in this regard. The $d=2$ case was solved completely in \cite{Bhatta:2016hpz}. The $d=1$ case can also be treated in full generality, which we present here.

\subsection{Complete $d=1$ analysis}
First we would like to compute the cap state for 1$d$ case and then the 1$d$ global blocks. We begin with the infinite dimensional matrix representations \cite{Jackiw:1990ka} of global conformal algebra $sl(2,\bf{R})$ for CFT$_1$:
\bea
L_{1}|h,n\rangle = \sqrt{n(2h+n-1)} ~ |h,n-1\rangle, &&  L_{-1}|h,n\rangle = \sqrt{(n+1)(2h+n)} ~|h,n+1\rangle \, , \cr
L_{0}|h,n\rangle &=& (h+n) \,|h,n\rangle
\eea
where $D = L_0$, $P = L_{-1}$ and $K= L_1$. The bulk is the ${\mathbb H}^2$ space whose tangent space rotation group is $SO(2)$. Therefore, the cap state $|h, \theta \rangle\!\rangle$ transforms as a 1-dimensional irrep of $SO(2)$:
\bea
(L_{1}-L_{-1})|h, \theta \rangle\!\rangle = \theta |h, \theta \rangle \! \rangle
\eea
The parameter $\theta$, a purely imaginary number, is related to the spin of the general bulk field -- we will elaborate further on this shortly. This equation can be solved for $|h, \theta \rangle \! \rangle$ as a linear combination of states in the module:
\bea
|h, \theta \rangle \!\rangle = \sum_{n=0}^{\infty} C_{n}|h,n\rangle \, ,
\eea
writing $C_{n}= \sqrt{\frac{\Gamma(2h)}{n! \, \Gamma(2h+n)}} ~ f_{n}$
with the $f_n$ satisfying the recursion relation
\bea
f_{n+1}=\theta \, f_{n}+~n~(2h+n-1) f_{n-1} \, .
\eea
It is not difficult to see that the $f_n$ are generated by $G(x)=\sum_{n=0}^{\infty}\frac{x^{n}}{n!} \, f_{n}$ where
\bea
G(x)=(1-x)^{-h-\tfrac{\theta}{2}}(1+x)^{-h+\tfrac{\theta}{2}} \, .
\eea
We find the coefficient of $\frac{x^n}{n!}$  in $G(x)$ to be:
\bea
f_{n}=(-1)^{n} \left(h-\frac{\theta}{2}\right)_{n} \, {}_{2}F_{1}\left(-n, \, h+\frac{\theta}{2}, \, -h+\frac{\theta}{2}-n+1;-1\right) \, .
\eea
Having obtained the expression for the most general cap state in $d=1$, we can repeat the rest of the exercises carried out in section 2 on these caps. Working with the coset element $g(x) = e^{\rho L_{0}} \, e^{-x L_{-1}}$ we can extract the leading terms in the large-$\rho$ limit of $\langle\!\langle h, \theta | g(x)|h,k\rangle$ and $\langle h,k | g^{-1}(y)|h, \theta \rangle\! \rangle$. With some further analysis we find the following simple answers in the $\rho\rightarrow \infty $ limit:
\bea
\label{1dlegs}
\lim_{\rho \rightarrow \infty} e^{\rho h} \langle\!\langle h, \theta | g(x)|h,k\rangle  
 &= &(-1)^{-h- \frac{\theta}{2}} \sqrt{\frac{\Gamma(2h+k)}{k!~ \Gamma(2h)}}  x^{-2h-k} \cr
\lim_{\rho \rightarrow \infty} e^{\rho h} \langle h,k | g^{-1}(y)|h, \theta \rangle\! \rangle &=& y^{k} \sqrt{\frac{\Gamma(2h+k)}{k! \, \Gamma(2h)}}
\eea
Notice that even though the general cap states depend on the spin-parameter $\theta$ the final expressions (\ref{1dlegs}) for the legs have essentially no dependence on it. For example, putting the legs together and performing the sum over $k$ gives 
\bea
\lim_{\rho \rightarrow \infty}  e^{2\rho h} \langle\!\langle h, \theta | g(x) g^{-1}(y)|h, \theta \rangle\!\rangle = (-1)^{-h- \tfrac{\theta}{2}} x^{-2h} \sum_{k=0}^{\infty} \frac{\Gamma(2h+k)}{k! ~~ \Gamma(2h)} \left(\frac{x}{y}\right)^k 
= (-1)^{-h- \tfrac{\theta}{2}} \frac{1}{(x-y)^{2 h}} \nonumber \\
\eea
A comparison of the $d=1$ legs here with the holomorphic part of the $d=2$ case of \cite{Bhatta:2016hpz} enables us to immediately write down the 1$d$ blocks by starting with the holomorphic parts of $d=2$ blocks and replacing $h \rightarrow \Delta$ and $z \rightarrow |x|$. It is evident that this will give rise to the 4-point block found above (\ref{1d4ptblock}), and higher-point ones to match with \cite{Qiao:2017xif, Gross:2017aos}.
\vskip .5cm
\noindent\underline{\bf The interpretation of $\theta$}
\vskip .5cm
To better understand the role of $\theta$ we must first look at the linearised bulk equations satisfied by the  the legs $\langle h,k | g^{-1}(x)| h, \theta \rangle\!\rangle$ and $\langle\!\langle h, \theta | g(x)|h,k\rangle$. To this end we first list the following identities \cite{Bhatta:2016hpz} satisfied by $g^{-1}(x)$
\bea
&& L_{0} ~ g^{-1}(x) = (-\partial_{\rho} +x \partial_{x})~g^{-1}(x),  ~~ L_{-1}~g^{-1}(x) = -\partial_{x}g^{-1}(x) \, ,\cr
&& L_{1}~g^{-1}(x)= (2 x \,  \partial_{\rho} -x^{2} \partial_{x} +e^{-\rho} \partial_{x})~g^{-1}(x) + g^{-1}(x) \, e^{-\rho} (L_{1}-L_{-1}) \, .
\eea
Using these we can easily compute the action of the $sl(2,\bf{R})$ Casimir operator $C_{2}$ on $g^{-1}(x)$
\bea
C_{2}&=& 2 L_{0}^{2} -  L_{1} L_{-1} -L_{-1} L_{1} \cr
C_{2} g^{-1}(x)& =& 2 \,(\partial^{2}_{\rho}+\partial_{\rho} +e^{-2 \rho} \partial_{x}^{2}) \, g^{-1}(x) +e^{-\rho} \partial_{x} g^{-1}(x)~(L_{1}-L_{-1})
\eea
Thus we see that the legs $\langle h,k | g^{-1}(x)| h, \theta \rangle\!\rangle$ satisfy the second order PDE:
\bea
\label{firstpde}
(\partial_{\rho}^{2} +\partial_{\rho} + e^{-2 \rho} \partial_{x}^{2}+ \theta \, e^{-\rho} \partial_{x}) \langle h,k | g^{-1}(x)| h, \theta \rangle\!\rangle = \Delta(\Delta-1)\langle h,k |g^{-1}(x)| h, \theta \rangle\!\rangle \, .
\eea
It is not difficult to see that the other legs $\langle\!\langle h, \theta | g(x)|h,k\rangle$ also satisfy the same equation. We would now like to interpret this equation as that of a bulk local field in the background $AdS_{2}$ geometry with metric $ds^2_{{\mathbb H}^2} = d\rho^2 + e^{2\rho} \, dx^2$.

Since the boundary isometry group is just ${\mathbb Z}_{2}$ we would expect the boundary conformal primary operators to be characterised by a scaling dimension $\Delta$ and a parity $\pm 1$. But any general bulk local field in two dimensions (once one trades off the spacetime indices for the tangent space ones) has to have only two parameters: the mass and the spin on which the bulk covariant derivative acts as
\bea
D_\mu \psi (x) = \partial_\mu \psi (x) + \frac{1}{2} \omega^{ab}_\mu L_{ab} \psi (x)
\eea
where $L_{ab}$ is the tangent space rotation generator in the representation of $\psi (x)$. Redefining the coordinates $z = e^{-\rho} +i x \, , ~ \bar{z} = e^{-\rho} - i x $
the metric of $AdS_2$ becomes $ds^{2} = \frac{4 dz d\bar{z}}{(z+\bar{z})^2}$.
For this geometry we have the following non-zero vielbeins, spin-connections and Christoffel connections:
\bea
e^{+} =\frac{ 2 dz}{(z+\bar{z})^2}, ~~ e^{-} = \frac{2 d\bar{z}}{(z+\bar{z})^2}, ~~
{\omega^{+}}_{+}=\frac{  dz- d\bar{z}}{(z+\bar{z})}= - {\omega^{-}}_{-}, ~~
\Gamma_{z \,z}^{z}=\frac{-2}{z+\bar{z}} =  \Gamma_{\bar{z}\,  \bar{z}}^{\bar{z}}
\eea
Since the tangent space is just ${\mathbb R}^2$, there is only one rotation generator $L_{+-}$, and we can take the field $\psi (x)$ to be an eigenstate of it with eigenvalue $i \, \theta$. Then it is easy to show that such a field satisfies the following
\bea
\label{secondpde}
(\square - m^2) \, \psi (x) &=& (\partial_{\rho}^{2} +\partial_{\rho} + e^{-2 \rho} \partial_{x}^{2}+\theta \,  e^{-\rho} \partial_{x}) \psi(x) -(m^2 + \frac{1}{8}\theta^2) \, \psi(x) =0
\eea
Comparing (\ref{firstpde}) and (\ref{secondpde}) we make the following identifications:
\bea
\Delta(\Delta-1)= m^{2}+\frac{\theta^2}{8} \, .
\eea
Therefore, we conclude that, when it is available, the parameter $\theta$ represents the spin of the bulk field.\footnote{Amusingly the same equation (\ref{secondpde}) arises for a complex scalar in $AdS_2$ minimally coupled to a background electric field preserving its isometries, and with strength $\theta$.}
\subsection{OWN{\it s} in more general $AdS_3$ geometries}
In \cite{Bhatta:2016hpz} we provided the computation of CPW of vacuum correlators in CFT$_2$ of primaries in general representations of the conformal algebra using the Wilson network prescription. Here we extend this result to include CPW of correlators in any (heavy) state, and to thermal correlators. This involves computing the OWN in appropriate locally $AdS_3$ geometries. Recall that in Fefferman-Graham gauge the most general solution to $AdS_3$ gravity \cite{Banados:1998gg} is
\bea
l^{-2} ds^2 &=& d\rho^2 + (dx_1^2 + dx_2^2) (e^{2\rho}  + T(z) \bar T(\bar{z}) \, e^{-2\rho}) \cr
&& ~~~~ + (T(z) + \bar T(\bar{z})) (dx_1^2 - dx_2^2) + 2i (T(z) - \bar T(\bar{z})) dx_1 dx_2 \nonumber
\eea
 When $- \infty < x_i < \infty$ -- it is a Euclidean locally $AdS_3$ geometry with boundary ${\mathbb R}^2$. For constant values of $T, \bar T \ge 0$ these are interpreted as BTZ black holes. When $-1/4 < T, \bar T <0 $ these represent heavy CFT states. We restrict to the constant $T, \bar T$ cases from now on. The relevant coset element is
\bea
g(x) &=& e^{\rho \, (L_0  + \bar L_0)} e^{-z \, (L_{-1} -   T \, L_1) } e^{-\bar z \, (\bar L_{-1} -  \bar{T}  \, \bar L_1) } \nonumber
\eea 
One can carry out the rest of the computations following \cite{Bhatta:2016hpz}. We find that the expressions for legs in the $\rho \rightarrow \infty$ are:
\bea
\lim_{\rho \rightarrow \infty} e^{\rho (h + \bar h)}
\langle \! \langle h, \bar h; j, m | g(x)|h, \bar h; k,\bar k \rangle & =& 
\tfrac{\lambda (-1)^{h - \bar h}}{\Gamma(2h) \sqrt{\Gamma(2h - 2\bar h)}} 
\left( \tfrac{\sqrt{T}}{\sinh (z  \sqrt{T})} \right)^{2h} 
\left( \tfrac{\sqrt{\bar {T}}}{\sinh (\bar z  \sqrt{\bar {T}})} \right)^{2\bar h}  \cr
&& ~~~~~ \times 
\sqrt{\tfrac{\Gamma(2h+k)}{k!} \tfrac{\Gamma(2\bar h + \bar k)}{{\bar k}!}} \left(\tfrac{\sqrt{T}}{\tanh (z \sqrt{T})}\right)^k
\left(\tfrac{\sqrt{\bar { T}}}{\tanh (\bar z \sqrt{\bar { T}})}\right)^{\bar k}  \cr \nonumber &&
\eea
\bea
\lim_{\rho \rightarrow \infty} e^{\rho (h + \bar h)} \langle h, \bar h; k, \bar k | g^{-1}(x)|h, \bar h; j,-j \rangle\!\rangle = \tfrac{\lambda}{\Gamma(2h) \sqrt{\Gamma(2j+1)}} \ 
 \tfrac{\left(\frac{\tanh(z \sqrt{{T}})}{\sqrt{{ T}}}\right)^{k}\left(\frac{\tanh(\bar z \sqrt{ \bar{{T}}})}{\sqrt{ \bar {{T}}}}\right)^{\bar k}}{\left(\cosh(z\sqrt { T})\right)^{2h}\left(\cosh(\bar z\sqrt{\bar{ {T}}})\right)^{2\bar h}} \sqrt{\tfrac{\Gamma(2h+k)\Gamma(2\bar h+\bar k)}{k! \ \ \bar k!}} \nonumber 
\eea
Putting them together for the 2-point function yields: 
\bea
\left\langle {\cal O}_{(h, \bar h)}(x_1) {\cal O}_{(h, \bar h)}(x_2) \right\rangle_{({T}, \bar{ T})} = \left(\tfrac{\sqrt{T}}{\sinh \left( (z_2 - z_1) \, \sqrt{T} \right)}\right)^{2h} \ \left(\tfrac{\sqrt{\bar {T}}}{\sinh \left( (\bar z_2 - \bar z_1) \sqrt{\bar {T}} \right) }\right)^{2 \bar h} \, .
\eea
which is the well known two-point function of a thermal CFT \cite{Gubser:1997cm, Datta:2011za} (see also \cite{Chen:2016uvu} and more recently \cite{Castro:2018srf}). The higher-point blocks can also be computed for these geometries \cite{ourunpub}.

\section{Discussion}
\label{sec6}
In this paper we have continued to develop further our prescription \cite{Bhatta:2016hpz} to compute the conformal partial waves of CFT correlation functions using the gravitational open Wilson network operators in the holographic dual gravity theories. In particular, we have demonstrated how to use gravitational Open Wilson Networks to compute 4-point scalar partial waves (both external and the exchanged operators being scalars) in any dimension. Our result for the scalar CPW are naturally given in the Gegenbauer polynomial basis. We have compared our results with the known answers wherever available and found complete agreement. Our methods also lead to a simpler proof of the recursion relation of \cite{SimmonsDuffin:2012uy} in even dimensional CFT{\it s}, and lead to analogous recursion relations for odd dimensions. 

The CPW for correlation functions of any primary with any exchanged operator has already been achieved in $d=2$ case in  \cite{Bhatta:2016hpz} and here in (in the simpler case of) $d=1$. It remains to generalise these computational techniques to obtain the CPWs for arbitrary representations in $d \ge 3$.\footnote{The generalisation of geodesic Witten diagrams to include some of the other representations was achieved in \cite{Nishida:2016vds, Dyer:2017zef, Tamaoka:2017jce, Nishida:2018opl}.} This involves finding the relevant cap states, and from there the relevant legs (conformal wave functions), and OPE modules etc. This work is in progress \cite{BRS3} and we hope to report on it in the near future. So far we have found the caps states for vectors, rank-2 antisymmetric and symmetric traceless tensor representations, and working on finding others. For those who may be interested, we present here the expressions of the cap states for the vector representation of the tangent space rotation algebra $so(d+1)$. This is constructed as a linear combination of the basis elements of the conformal module over a vector primary state and it is given by:
\bea
\label{vectorcap}
|\phi_\Delta, \alpha \rangle\!\rangle &=& \sum_{n=0}^\infty  (P_\gamma P^\gamma)^n \, \sum_{\beta=1}^d A^{(n)}_{\alpha\beta}(\Delta, d) \, |\phi_\Delta, \beta \rangle + \sum_{n=0}^\infty (P_\gamma P^\gamma)^n B^{(n)} (\Delta, d) \, P_\alpha \, \sum_{\gamma =1}^d P_\gamma |\phi_\Delta, \gamma \rangle \cr
|\phi_\Delta, d+1 \rangle\!\rangle &=& \sum_{n=0}^\infty C^{(n)} (\Delta, d) \, (P_\gamma P^\gamma)^n \sum_{\beta =1}^d P_\beta |\phi_\Delta, \beta \rangle 
\eea
with  $A^{(n)} = C^{(n)} (\Delta_\phi-1)$ and $B^{(n)} = \frac{ C^{(n)}}{2}+ 2(n+1)C^{(n+1)} \left(\Delta_\phi + n - \frac{d}{2} +2 \right)$.
\vskip .5cm
Of course one would like to see if our method gives answers in forms more amenable to potential applications, such as in the bootstrap approach towards the classification of CFT{\it s}. Since our answers are in Gegenbauer polynomial basis it is possible that they may be found more suitable -- as working with this basis is much simpler (as we have seen in section 4, for example).

An interesting set of future directions should include exploring the role of Weight shifting  operators \cite{Costa:2011dw, Karateev:2017jgd, Costa:2018mcg} in our formalism.

It may be of interest to compute objects similar to our OWN{\it s} in both flat and de Sitter gravity theories. Such diagrams could provide a basis of partial waves for S-matrices for scattering problems in these spaces. 

Another possible generalisation should involve inclusion of boundaries and defects to the CFT \cite{McAvity:1995zd, Nakayama:2016cim, Gadde:2016fbj} in the formalism considered.
 
We hope that this program will naturally lend itself to answering dynamical questions as well in CFT{\it s}. 

\section*{Acknowledgements}
We would like to thank ICTS, Bengaluru for hospitality during the final stages of this work, and the participants of the workshop ``AdS/CFT @ 20 and Beyond" for discussions. AB would like to thank IoP, Bhubaneswar for hospitality while the work was in progress.
\appendix
\section{CGC required for scalar CPW}
\label{appA}
Here we record results of CGC of the representations considered in the text for the algebra $so(1, d+1)$. These are needed to compute the CPW for higher-point functions from OWNs. Before we present the detailed derivation of the CGC from the 3-point functions we collect a few facts about the irreducible representations of $so(d)$ which we will use in the extraction of the Clebsch-Gordan coefficients.

A finite dimensional irreducible representation of $so(d)$ is uniquely defined by its highest weight $[\mu_{1},\mu_{2},...\mu_{k}]$
with 
\bea
&&\mu_1 \geq \mu_2 \geq \cdots \geq \mu_{k-1} \geq |\mu_{k}| ~~~~~~ \text{for $d=2k$} \cr
&&\mu_1 \geq \mu_2 \geq \cdots \geq \mu_{k-1} \geq \mu_{k} \geq 0 ~~\text{for $d=2k+1$}
\eea
The components $\mu_{i}$ are either simultaneously integers (tensorial representations) or half-integers (spinorial representations). We only consider symmetric traceless representations of $so(d)$ as these are the only relevant ones for the scalar CGC of $so(d+1,1)$. These could be  represented on the Hilbert space $H$ of square integrable function on $S^{d-1}$. The Hilbert space can be decomposed into an orthogonal sum of subspaces $H^{l}$ of homogenous polynomials of degree $l$ in $d$ variables. We introduce a complete orthonormal basis ${|l, {\bf M} \rangle}$ on $H^{l}$, where ${\bf M} =(m_{d-2},m_{d-3},...,m_2,m_1)$ label these basis states provided they fulfil:
\bea
l=m_{d-1} \geq m_{d-2}\geq \cdots \geq m_2 \geq |m_1| ~~~~~ m_1 \in {\mathbb Z}  ~~~ m_i \in {\mathbb Z}_{>0} ~~i\geq 2
\eea
The dimension of the space $H$ is $d_{l}= (2l+d-2) \frac{(l+d-3)!}{l!(d-2)!}$ - the number of independent components of a general symmetric traceless tensor of rank $l$ in $d$ dimensions. The matrix elements of the representation $D^{l}$ read:
\bea
D^{l}_{{\bf M}\, {\bf M}'}(g) = \langle l, {\bf M}|D^{l}(g) |l,{\bf M}'\rangle
\eea
In particular,
\bea
D^{l}_{{\bf M}\, {\bf 0}}(g) = \frac{1}{\sqrt{d_l}}~N^{d}_{l\, {\bf M}} \prod_{k=1}^{d-2} C_{m_{k+1}-m_{k}}^{m_{k}+k/2} \,\cos (\Phi_{k+1}) ~ \sin^{m_k}(\Phi_{k+1}) \, e^{i m_{1} \Phi_{1}}
\eea
where $N^{d}_{l\, {\bf M}}$ is the normalisation w.r.t the Haar measure on $so(d)$, $C_{\lambda}^{n}(z)$ are the Gegenbauer polynomials. The angles $0 \leq \Phi_{1}\leq 2 \pi$ and $0 \leq \Phi_{i}\leq \pi$ for $i\neq 1$ can be identified with  the Euler angles of a rotation $g$ which maps the north pole $a=(0,\cdots,0,1) \in {\mathbb R}^d$ to an arbitrary point on $S^{d-1}$. Then the hyperspherical harmonics on $S^{d-1}$ are defined as follows:
\bea
|e\rangle = D^{l}(g) |a\rangle, ~~ Y_{l \, {\bf M}}(e) = \langle e| l, \, {\bf M} \rangle, ~~  \langle a| l,\,{\bf M}\rangle = \sqrt{\frac{d_{l}}{V_{d}}} ~ \delta_{{\bf M} \, {\bf 0}}
\eea
where $V_{d}= \frac{2 \pi^{d/2}}{\Gamma(d/2)}$ is the volume of unit $S^{d-1}$ sphere. Therefore, we get
\bea
Y_{l \, {\bf M}}(e) &= &\sqrt{\frac{d_{l}}{V_{d}}} ~~ D^{l\,*}_{{\bf M} \, {\bf 0}}(g) 
\eea
We finally list the following properties of hyperspherical harmonics which can be easily derived using the definitions given above:
\begin{enumerate}
\item $ Y^{*}_{l\, {\bf M}}(e) = (-1)^{m_1} Y_{l \, \bar{{\bf M}}}(e) $ ~~~where ~~~ $\bar{{\bf M}}=(m_{d-2},\cdots,m_2,-m_1)$.
\item $
Y_{l_1\, {\bf M}_1}(e)Y_{l_2\, {\bf M}_2}(e) =  \sum_{l_3, {\bf M}_3} \left( {l_1 \, l_2 \, l_3}\atop{{\bf M}_1\, {\bf M}_2 \,{\bf M}_3} \right)  \left( {l_1 \, l_2 \, l_3}\atop{{\bf 0} \, {\bf 0} \,{\bf 0}} \right) Y^{*}_{l_3\,{\bf M}_3}(e)$

\item $\left( {l_1 \, l_2 \, l_3}\atop{0 \, 0 \,0}\right) = 0$ unless $l_1+l_2+l_3$ is an even integer and $l_3= |l_1-l_2|,\cdots,l_1+l_2$.

\item
$\left( {l \, l' \, 0}\atop{{\bf M} \, {\bf M}' \,0}\right) = \frac{(-1)^{l-m1}}{\sqrt{d_{l}}}\delta_{l \, l'}\delta_{{\bf M}\,{\bf M}'}$

\item 
$ \sum_{\{{\bf m}_i\}} (-1)^{({\bf m}_2)_1} \,\left({l_1\,l_3 \,L_2}\atop{{\bf m}_1\, {\bf m}_3\,{\bf M}_2}\right)\left({l_1\,l_2 \,L_3}\atop{\bar{{\bf m}}_1\, {\bf m}_2\,{\bf M}_3}\right)\left({l_2\,l_3 \,L_1}\atop{\bar{{\bf m}}_2 \,\bar{{\bf m}}_3\,{\bf M}_1}\right)= (-1)^{l_2+L_2-L_3} \,\left({ L_3\,L_1\,L_2}\atop{\bar{{\bf M}}_3 \,\bar{{\bf M}}_1 \,{\bf M}_2 } \right)\, \left\{\begin{matrix}
 	L_1 & L_2 & L_3 \\
 	l_1 & l_2 & l_3
 \end{matrix}
 \right\} $
 \end{enumerate}

\vskip .5cm
\noindent\underline{\bf $so(1, d+1)$ CGC for Scalar irreps}
\vskip .5cm
We extract CG coefficients for $so(1,d+1)$ for three scalars from the three-point function for scalar primary operators amputating the out-going and in-going legs we had found earlier. The out-going and in-going legs takes the following forms respectively
\bea
\lim_{\rho\rightarrow \infty} e^{\rho\Delta} \langle\!\langle \Delta |g(x)|\Delta ; \{l,{\bf m}, s\}\rangle &=& \frac{ 2^{l+2s}}{A_{l,s}} \frac{\Gamma(\Delta+s+l)\,\Gamma(\Delta+s- \mu)}{\Gamma(\Delta)\,\Gamma(\Delta- \mu)} \,  M^{l}_{{\bf m}}({\bf x}) \,  (x^2)^{-\Delta-l-s} \cr 
&& \cr 
&=& \left[ \frac{(\Delta)_{l+s} (\Delta- \mu)_{s}}{(\mu+1)_{l+s}s!}\right]^{1/2} \,  (x^2)^{-\Delta-l-s}\, M^{l^{}}_{{\bf m}}({\bf x})\cr 
&&
\eea
and
\bea
\lim_{\rho\rightarrow \infty} e^{\rho\Delta} \langle \Delta; \{l,{\bf m},s\}|g^{-1}(y)|\Delta \rangle\!\rangle &=& \frac{A_{l,s}}{2^{l+2s}}\, \frac{(y^2)^{s}}{(s)!\, \Gamma(l+s+3/2)}  \,M^{l^{\,*}}_{{\bf m}}({\bf y}) \cr 
&& \cr 
&=&\left[ \frac{(\Delta)_{l+s} (\Delta- \mu)_{s}}{(\mu+1)_{l+s}s!}\right]^{1/2} \, (y^2)^{s} \, M^{l^{\,*}}_{{\bf m}}({\bf y}) \cr 
&&
\eea
The 3-point function of the scalar primary operators with conformal dimensions $\Delta_1, \Delta_2$ and $\Delta_3$
\bea
\frac{1}{|x_2-x_1|^{\Delta_1+\Delta_2-\Delta_3}\, |x_3-x_2|^{\Delta_2+\Delta_3-\Delta_1}\,|x_3-x_1|^{\Delta_1+\Delta_3-\Delta_2}}
\eea
can be expanded as
\bea
 && \left(4 \pi^{ \,d/2}\right)^3 \prod_{i=1}^{3} \sum_{l_i=0}^{\infty} \sum_{s_i=0}^{\infty} \sum_{{\bf m}_i} \tfrac{(\Delta_{12}/2)_{l_1+s_1}(\Delta_{12}/2-\mu)_{s1}}{(\mu+1)_{l_1+s_1} s_1!}  \tfrac{(\Delta_{23}/2)_{l_2+s_2}(\Delta_{23}/2-\mu)_{s2}}{(\mu+1)_{l_2+s_2} s_2!}  \tfrac{(\Delta_{31}/2)_{l_3+s_3}(\Delta_{31}/2-\mu)_{s3}}{(\mu+1)_{l_3+s_3} s_3!}\cr 
 && \cr 
 && ~~~~~~~~~~\times (x^2)^{-\Delta_1-l_1-l_3-s_1-s_3} \, (y^2)^{-\Delta_{23}/2-l_2+s_1-s_2} \, (z^2)^{s_2+s_3} \cr
 && ~~~~~~~~~~\times M^{l_1}_{{\bf m}_1}({\bf x})\, M^{l_3}_{{\bf m}_3}({\bf x}) \, M^{l_1^{\,*}}_{{\bf m}_1}({\bf y}) \, M^{l_2}_{{\bf m}_2}({\bf y})\,M^{l_2^{\,*}}_{{\bf m}_2}({\bf z}) \, M^{l_3^{\,*}}_{{\bf m}_3}({\bf z}) \cr 
 &&
\eea
where
\bea 
\Delta_{12} \equiv \Delta_1 +\Delta_2 -\Delta_3, ~~~
\Delta_{23} \equiv \Delta_2 +\Delta_3 -\Delta_1, ~~~
\Delta_{31} \equiv \Delta_3 +\Delta_1 -\Delta_2 
\eea
We use the following identities: 
\bea
\label{eq:sp_harmonics_decomp}
M^{l}_{{\bf m}}({\bf x}) M^{l'}_{{\bf m}'}({\bf x}) = \sum_{L,M} \left( {l \, l' \, L}\atop{{\bf m} \,{\bf m}' \,{\bf M}} \right)  \left( {l \, l' \, L}\atop{{\bf 0} \, {\bf 0} \,{\bf 0}} \right) (x^{2})^{\frac{l+l'-L}{2}} M^{L^{\,*}}_{{\bf M}}({\bf x}) 
\eea
\bea
M^{l^{\,*}}_{{\bf m}}(\bf {x}) =(-1)^{m_{1}}  M^{l}_{ \bar{m} } (\bf{ x})
\eea
where,
$\bar{{\bf m}}=(m_{n-2},...,m_{2},-m_1)$ to rewrite the product of spherical harmonics in the summand as
\bea
 && M^{l_1}_{{\bf m}_1}({\bf x}) M^{l_3}_{{\bf m}_3}({\bf x}) M^{l_1^{\,*}}_{{\bf m}_1}({\bf y}) M^{l_2}_{{\bf m}_2}({\bf y})M^{l_2^{\,*}}_{{\bf m}_2}({\bf z})M^{l_3^{\,*}}_{{\bf m}_3}({\bf z}) \cr \cr
&=& (-1)^{m_1+m_2+m_3}\,M^{l_1}_{{\bf m}_1}({\bf x}) M^{l_3}_{{\bf m}_3}({\bf x}) M^{l_1}_{\bar{{\bf m}_1}}({\bf y})M^{l_2}_{{\bf m}_2}({\bf y})M^{l_2}_{\bar{{\bf m}_2}}({\bf z}) M^{l_3}_{\bar{{\bf m}_3}}({\bf z})  \cr 
 &=& (-1)^{m_2} \prod_{i=1}^{3} \sum_{L_i,M_i} \,(-1)^{M_2}
 \left({l_1\,l_3 \,L_2}\atop{{\bf 0}\,{\bf 0}\,{\bf 0}}\right)\left({l_1\,l_2 \,L_3}\atop{{\bf 0}\,{\bf 0}\,{\bf 0}}\right)\left({l_2\,l_3 \,L_1}\atop{{\bf 0}\,{\bf 0}\,{\bf 0}}\right)\cr 
 && ~~\times\left({l_1\,l_3 \,L_2}\atop{{\bf m}_1\,{\bf m}_3\,{\bf M}_2}\right)\left({l_1\,l_2 \,L_3}\atop{\bar{{\bf m}_1}\,{\bf m}_2\,{\bf M}_3}\right)\left({l_2\,l_3 \,L_1}\atop{\bar{{\bf m}_2}\,\bar{{\bf m}_3}\,{\bf M}_1}\right) \cr 
 && ~~\times (x^2)^{\frac{l_1+l_3-L_2}{2}} \, (y^2)^{\frac{l_1+l_2-L_3}{2}} \,(z^2)^{\frac{l_2+l_3-L_1}{2}} \, M^{L_2}_{{\bf M}_2} ({\bf x}) \, M^{L_3}_{{\bf M}_3} ({\bf y}) \, M^{L_1}_{{\bf M}_1} ({\bf z})
\eea
Inserting the above relation in the summand and performing the $\{m_i\}$ summations we get
\bea
 && \sum_{\{{\bf m}_i\}} (-1)^{m_2} \,\left({l_1\,l_3 \,L_2}\atop{{\bf m}_1\,{\bf m}_3\,{\bf M}_2}\right)\left({l_1\,l_2 \,L_3}\atop{\bar{{\bf m}_1}\,{\bf m}_2\,{\bf M}_3}\right)\left({l_2\,l_3 \,L_1}\atop{\bar{{\bf m}_2}\,\bar{{\bf m}_3}\,{\bf M}_1}\right) \cr 
 && \cr 
 &&~ = (-1)^{l_2+L_2-L_3} \,\left({ L_3\,L_1\,L_2}\atop{\bar{{\bf M}}_3 \,\bar{{\bf M}}_1 \,{\bf M}_2 } \right) \left\{ \begin{matrix}
 	      l_2 & l_1 & L_3 \\
 	      L_2 & L_1 & l_3
 	    \end{matrix}
 	 \right\} \cr 
 && \cr 
 &&~ = (-1)^{l_2+L_2-L_3} \,\left({ L_3\,L_1\,L_2}\atop{\bar{{\bf M}}_3 \,\bar{{\bf M}}_1 \,{\bf M}_2 } \right)\, \left\{ \begin{matrix}
 	L_1 & L_2 & L_3 \\
 	l_1 & l_2 & l_3
 \end{matrix}
 \right\}
\eea
Now the three-point function takes the form
\bea
&&\left(4 \pi^{ \,d/2}\right)^3 \prod_{i=1}^{3} \sum_{l_i=0}^{\infty} \sum_{s_i=0}^{\infty} \sum_{{\bf m}_i} \tfrac{(\Delta_{12}/2)_{l_1+s_1}(\Delta_{12}/2-\mu)_{s1}}{(\mu+1)_{l_1+s_1} s_1!}  \tfrac{(\Delta_{23}/2)_{l_2+s_2}(\Delta_{23}/2-\mu)_{s2}}{(\mu+1)_{l_2+s_2} s_2!}  \tfrac{(\Delta_{31}/2)_{l_3+s_3}(\Delta_{31}/2-\mu)_{s3}}{(\mu+1)_{l_3+s_3} s_3!} \cr 
 && \prod_{i=1}^{3} \sum_{L_{i}, {\bf M}_{i}} (x^2)^{-\Delta_1-l_1-l_3-s_1-s_3+\frac{l_1+l_3-L_2}{2}} \, (y^2)^{-\Delta_{23}/2-l_2+s_1-s_2+\frac{l_1+l_2-L_3}{2}} \, (z^2)^{s_2+s_3+\frac{l_2+l_3-L_1}{2}} \cr 
 && (-1)^{l_2+L_2-L_3+M_2} M^{L_2}_{{\bf M}_2} ({\bf x}) \, M^{L_3}_{{\bf M}_3} ({\bf y}) \, M^{L_1}_{{\bf M}_1} ({\bf z})  \cr
 && \left({l_1\,l_3 \,L_2}\atop{{\bf 0}\,{\bf 0}\,{\bf 0}}\right)\left({l_1\,l_2 \,L_3}\atop{{\bf 0}\,{\bf 0}\,{\bf 0}}\right)\left({l_2\,l_3 \,L_1}\atop{{\bf 0}\,{\bf 0}\,{\bf 0}}\right)\left({ L_3\,L_1\,L_2}\atop{\bar{{\bf M}}_3 \,\bar{{\bf M}}_1 {\bf M}_2 } \right)\, \left\{ \begin{matrix}
 	L_1 & L_2 & L_3 \\
 	l_1 & l_2 & l_3
 \end{matrix}
 \right\} \nonumber
\eea
which can also be written as
\bea
&&\left(4 \pi^{ \,d/2}\right)^3 \prod_{i=1}^{3} \sum_{l_i=0}^{\infty} \sum_{s_i=0}^{\infty} \sum_{{\bf m}_i} \tfrac{(\Delta_{12}/2)_{l_1+s_1}(\Delta_{12}/2-\mu)_{s1}}{(\mu+1)_{l_1+s_1} s_1!}  \tfrac{(\Delta_{23}/2)_{l_2+s_2}(\Delta_{23}/2-\mu)_{s2}}{(\mu+1)_{l_2+s_2} s_2!}  \tfrac{(\Delta_{31}/2)_{l_3+s_3}(\Delta_{31}/2-\mu)_{s3}}{(\mu+1)_{l_3+s_3} s_3!} \cr 
 && \prod_{i=1}^{3} \sum_{L_{i}, M_{i}} (x^2)^{-\Delta_1-l_1-l_3-s_1-s_3+\frac{l_1+l_3-L_2}{2}} \, (y^2)^{-\Delta_{23}/2-l_2+s_1-s_2+\frac{l_1+l_2-L_3}{2}} \, (z^2)^{s_2+s_3+\frac{l_2+l_3-L_1}{2}} \cr 
 && (-1)^{l_2+L_2-L_3} M^{L_2}_{\bar{{\bf M}}_2} ({\bf x}) \, M^{L^{\;*}_3}_{\bar{{\bf M}}_3} ({\bf y}) \, M^{L^{\,*}_1}_{\bar{{\bf M}}_1} ({\bf z})  \cr
 && \left({l_1\,l_3 \,L_2}\atop{{\bf 0}\,{\bf 0}\,{\bf 0}}\right)\left({l_1\,l_2 \,L_3}\atop{{\bf 0}\,{\bf 0}\,{\bf 0}}\right)\left({l_2\,l_3 \,L_1}\atop{{\bf 0}\,{\bf 0}\,{\bf 0}}\right)\left({ L_3\,L_1\,L_2}\atop{\bar{{\bf M}}_3 \,\bar{{\bf M}}_1 {\bf M}_2 } \right)\, \left\{ \begin{matrix}
 	L_1 & L_2 & L_3 \\
 	l_1 & l_2 & l_3
 \end{matrix}
 \right\} \nonumber
\eea
Note that the $so(d)$ $3\,j$- coefficients $\left({l \,l' \,L}\atop{0\,0\,0}\right)$ is non-vanishing only when $l+l'-L$ is even integer. This suggests to change the following variables as
\bea
 l_1+l_3 = 2K_2 +L_2, ~~~
 l_1+l_2 = 2K_3 +L_3, ~~~
 l_2+l_3 = 2K_1 +L_1
\eea
i.e.
\bea
 l_1 &=& \frac{L_2+L_3-L_1}{2} + K_2+K_3-K_1 \cr
 l_2 &=& \frac{L_3+L_1-L_2}{2} + K_3+K_1-K_2 \cr
 l_3 &=& \frac{L_1+L_2-L_3}{2} + K_1+K_2-K_3 
\eea
Then the powers of $x^2,\, y^2,\, z^2$ becomes (excluding the powers within the spherical harmonics)
\bea
 (x^2)^{-\Delta_1-L_2-K_2-s_1-s_3} \, (y^2)^{-\frac{\Delta_{23}}{2}-\frac{L_3+L_1-L_2}{2}+s_1-s_2-K_1+K_2} \, (z^2)^{s_2+s_3+K_1}
\eea
respectively. Comparing with the legs we want to amputate from the three-point function we make the following change of variables in the summand
\bea
 K_2+s_1+s_3 &=& S_2 \\
 K_1+s_2+s_3 &=& S_1 \\
 -\frac{\Delta_{23}}{2}-\frac{L_3+L_1-L_2}{2}+s_1-s_2-K_1+K_2 &=& S_3
\eea
The last one of the above relations impose the following selection rule
\bea
 (\Delta_2 +L_1+2S_1) + (\Delta_3+L_3+2S_3) = (\Delta_1+L_2+2S_2)
\eea
So $S_3$ is not an independent variable and $s_3$ is undetermined in terms of new variables. We call it $s_3 =S$. In terms of the new variables the three-point function becomes
\bea
 && \frac{\left(4\pi^{d/2}\right)^3}{\Gamma(
 	\Delta_{12}/2)\Gamma(
 	\Delta_{23}/2)\Gamma(
 	\Delta_{31}/2)\Gamma(
 	\Delta_{12}/2-\mu)\Gamma(
 	\Delta_{23}/2-\mu)\Gamma(
 	\Delta_{31}/2-\mu)} \cr
	&& \times \prod_{i=1}^{3} \sum_{L_i=0}^{\infty} \sum_{S_i=0}^{\infty} \sum_{M_i}
 \delta (\Delta_2+L_1+2S_1+\Delta_3+L_3+2S_3-\Delta_1-L_2-2S_2) \cr 
 && \cr 
 && \sum_{K_3=0}^{\infty}\sum_{K_1=0}^{S_1}\sum_{K_2=0}^{S_2}\sum_{S=0}^{\text{min}(S_2-K_2, S_1-K_1)}\tfrac{\Gamma(\Delta_{2}+L_3+S_1+S_3+K_3-S-K_1)\,\Gamma(\Delta_{12}/2+S_2-K_2-S-\mu)}{\Gamma(\frac{L_2+L_3-L_1}{2} +K_3-K_1+S_2-S+d/2) \,(S_2-K_2-S)!} \cr 
 && \cr 
 &&~~~~ ~~~\times \tfrac{\Gamma(K_3-K_2+S_2-S_3-S)\,\Gamma(\Delta_{23}/2+S_1-K_1-S-\mu)}{\Gamma(\frac{L_3+L_1-L_2}{2}+K_3-K_2+S_1-S+d/2)\,(S_1-K_1-S)!} \cr 
 && \cr  
 &&\times \tfrac{\Gamma(\Delta_{3}+L_1+S_1+S_3-S_2+K_1+K_2-K_3+S)\,\Gamma(\Delta_{31}/2+S-\mu)}{\Gamma(\frac{L_1+L_2-L_3}{2}+K_1+K_2-K_3+S+d/2)\,S!} \times  (-1)^{\tfrac{L_1+L_2-L_3}{2}+K_3+K_1-K_2} \cr 
 && \cr 
 && \times \left({\frac{L_2+L_3-L_1}{2} + K_2+K_3-K_1, \, \frac{L_1+L_2-L_3}{2} + K_1+K_2-K_3 ,\,L_2 }\atop{0 ~~~~~~~~ \,  0~~~~~~~~~ \,0}  \right) \cr 
 && \cr 
 && \times \left({ \frac{L_2+L_3-L_1}{2} + K_2+K_3-K_1,\, \frac{L_3+L_1-L_2}{2} + K_3+K_1-K_2,\,L_3} \atop{0 ~~~~~~~~ \,  0~~~~~~~~~ \,0}  \right)  \cr 
 && \cr 
 && \times \left({ \frac{L_3+L_1-L_2}{2} + K_3+K_1-K_2,\, \frac{L_1+L_2-L_3}{2} + K_1+K_2-K_3,\,L_1}\atop{0 ~~~~~~~~ \,  0~~~~~~~~~ \,0}  \right)  \cr 
 && \cr 
 && \times \left\{ \begin{matrix}
 	L_1 &  L_2 &  L_3 \\
 	\frac{L_2+L_3-L_1}{2} + K_2+K_3-K_1 ~~ & ~~ \frac{L_3+L_1-L_2}{2} + K_3+K_1-K_2 ~~ & ~~ \frac{L_1+L_2-L_3}{2} + K_1+K_2-K_3
 \end{matrix}
 \right\} \cr 
 && \cr 
 && \times \left({ L_3\,L_1\,L_2}\atop{{\bf M}_3 \,{\bf M}_1 \,{\bf M}_2 } \right)\, (x^2)^{-\Delta_1-L_2-S_2} (y^2)^{S_1} (z^2)^{S_3} \, M^{L_2^{\, *}}_{-{\bf M}_2} ({\bf x}) \, M^{L_3}_{-{\bf M}_3} ({\bf y}) \, M^{L_1}_{-{\bf M}_1} ({\bf z}) 
\eea
where we have arranged the order as well as the limits of the summations appropriately. According to our prescription the three-point function can be recovered as
\bea
 && \prod_{i=1}^{3} \sum_{L_i=0}^{\infty} \sum_{S_i=0}^{\infty} \sum_{{\bf M}_i} \langle\!\langle \Delta_1 |g(x)|\Delta_1 ; \{L_2,{\bf M}_2, S_2\}\rangle \, \langle \Delta_2; \{L_1,{\bf M}_1,S_1\}|g^{-1}(y)|\Delta_2 \rangle\!\rangle \cr 
 && \qquad \qquad \qquad \qquad \qquad \langle \Delta_3; \{L_3,{\bf M}_3,S_3\}|g^{-1}(z)|\Delta_3 \rangle\!\rangle \, C^{(\Delta_2,S_1),\, (\Delta_3,S_3);\, (\Delta_1,S_2)}_{(L_1,{\bf M}_1),\, (L_3,{\bf M}_3);\, (L_2,{\bf M}_2)}
\eea
where $C^{(\Delta_2,S_1),\, (\Delta_3,S_3);\, (\Delta_1,S_2)}_{(L_1,{\bf M}_1),\, (L_3,{\bf M}_3);\, (L_2,{\bf M}_2)}$ is $so(1,d+1)$ CG coefficient. Comparing above with the three-point function we write
\bea
 && \cr 
 && C^{(\Delta_2,S_1),\, (\Delta_3,S_3);\, (\Delta_1,S_2)}_{(L_1,{\bf M}_1),\, (L_3,{\bf M}_3);\, (L_2,{\bf M}_2)} \cr 
 && \cr 
 && = \frac{\left(4\pi^{d/2} \right)^{3} \,\delta (\Delta_2+L_1+2S_1+\Delta_3+L_3+2S_3-\Delta_1-L_2-2S_2)}{\Gamma(
 	\Delta_{12}/2)\Gamma(
 	\Delta_{23}/2)\Gamma(
 	\Delta_{31}/2)\Gamma(
 	\Delta_{12}/2-\mu)\Gamma(
 	\Delta_{23}/2-\mu)\Gamma(
 	\Delta_{31}/2-\mu)} \cr
 && ~\times \left[ \frac{\Gamma(\Delta_1+L_2+S_2)\, \Gamma(\Delta_1+S_2-\mu)}{\Gamma(\Delta_1)\Gamma(\Delta_1-\mu)\Gamma(L_2+S_2+d/2) S_2!}\right]^{1/2} \left[ \frac{\Gamma(\Delta_2+L_1+S_1)\, \Gamma(\Delta_2+S_1-\mu)}{\Gamma(\Delta_2)\Gamma(\Delta_2-\mu)\Gamma(L_1+S_1+d/2) S_1!}\right]^{1/2} \cr 
 && \cr 
 && ~\times \left[ \frac{\Gamma(\Delta_3+L_3+S_3)\, \Gamma(\Delta_3+S_3-\mu)}{\Gamma(\Delta_3)\Gamma(\Delta_3-\mu)\Gamma(L_3+S_3+d/2) S_3!}\right]^{1/2}  \cr && \times \sum_{K_3=0}^{\infty}\sum_{K_1=0}^{S_1}\sum_{K_2=0}^{S_2}\sum_{S=0}^{min(S_2-K_2, S_1-K_1)} (-1)^{\tfrac{L_1+L_2-L_3}{2}+K_3+K_1-K_2} ~~  \cr
&&  \times \tfrac{\Gamma(\Delta_{2}+L_3+S_1+S_3+K_3-S-K_1)\,\Gamma(\Delta_{12}/2+S_2-K_2-S-\mu)}{\Gamma(\frac{L_2+L_3-L_1}{2} +K_3-K_1+S_2-S+d/2) \,(S_2-K_2-S)!} \cr 
 && \cr 
 &&\times \tfrac{\Gamma(K_3-K_2+S_2-S_3-S)\,\Gamma(\Delta_{23}/2+S_1-K_1-S-\mu)}{\Gamma(\frac{L_3+L_1-L_2}{2}+K_3-K_2+S_1-S+d/2)\,(S_1-K_1-S)!} \cr 
 && \cr
 && \times \tfrac{\Gamma(\Delta_{3}+L_1+S_1+S_3-S_2+K_1+K_2-K_3+S)\,\Gamma(\Delta_{31}/2+S-\mu)}{\Gamma(\frac{L_1+L_2-L_3}{2}+K_1+K_2-K_3+S+d/2)\,S!} \cr
 && \cr 
 && \times \left({\frac{L_2+L_3-L_1}{2} + K_2+K_3-K_1, \, \frac{L_1+L_2-L_3}{2} + K_1+K_2-K_3 ,\,L_2 }\atop{0 ~~~~~~~~ \,  0~~~~~~~~~ \,0}  \right) \cr 
 && \cr 
 && \times \left({ \frac{L_2+L_3-L_1}{2} + K_2+K_3-K_1,\, \frac{L_3+L_1-L_2}{2} + K_3+K_1-K_2,\,L_3} \atop{0 ~~~~~~~~ \,  0~~~~~~~~~ \,0}  \right)  \cr 
 && \cr 
 && \times \left({ \frac{L_3+L_1-L_2}{2} + K_3+K_1-K_2,\, \frac{L_1+L_2-L_3}{2} + K_1+K_2-K_3,\,L_1}\atop{0 ~~~~~~~~ \,  0~~~~~~~~~ \,0}  \right)  \cr 
 && \cr 
 && \times \left\{ \begin{matrix}
 	L_1 &  L_2 &  L_3 \\
 	\frac{L_2+L_3-L_1}{2} + K_2+K_3-K_1 ~~ & ~~ \frac{L_3+L_1-L_2}{2} + K_3+K_1-K_2 ~~ & ~~ \frac{L_1+L_2-L_3}{2} + K_1+K_2-K_3
 \end{matrix}
 \right\} \cr 
 && \cr 
 && \times \left({ L_3\,L_1\,L_2}\atop{{\bf M}_3 \,{\bf M}_1 \,{\bf M}_2 } \right) 
\eea
\section{Manipulation of the $d$-dimensional result}
\label{appB}
The four-point block of a scalar correlation function in general dimensions in our method takes the following form:
\bea
(x^2)^{\frac{1}{2}(\Delta - \Delta_3-\Delta_4)} 
&& \sum_{l,s} \Gamma(\alpha+l+s) \Gamma(\beta+l+s) 
\,  \Gamma(\alpha+s- \mu)\Gamma(\beta+s-\mu) \cr
&& ~~~  \frac{\left(l+\mu \right)}{s! \, \Gamma(l+s+1+\mu) \Gamma(\Delta + l +s) \Gamma(\Delta+s-\mu)} \, (x^2)^s \, {\cal C}^\mu_l ({\bf x} \cdot {\bf u})
\eea
where $\mu = \frac{d-2}{2}$.
One of the questions we have to address is how our computations match with those known in the literature. There is a famous expression for the conformal blocks in any dimension in terms the cross ratios $u, v$ as found by Dolan and Osborn. We now prove the following identity towards establishing the equivalence between our answers and theirs. 
\bea
\sum_{l,s=0}^\infty \tfrac{\left(\frac{\Delta - \Delta_{12}}{2}\right)_{l+s} \left(\frac{\Delta + \Delta_{34}}{2}\right)_{l+s}}{(\Delta)_{l+s}}   
\tfrac{\left(\frac{\Delta - \Delta_{12}}{2} - \mu\right)_{s} \left(\frac{\Delta  +\Delta_{34}}{2} - \mu\right)_{s}}{(\Delta -\mu)_{s}} \tfrac{1+\frac{l}{\mu}}{s! (\mu+1)_{l+s} } (z \bar z)^{s+ \frac{l}{2}} C_l^\mu (\tfrac{z+ \bar z}{2 \sqrt{z \bar z}})
\eea
is equal to
\bea
\sum_{r,q=0}^\infty \tfrac{\left(\frac{\Delta + \Delta_{12}}{2}\right)_{r} \left(\frac{\Delta - \Delta_{12}}{2} \right)_{r+q} \left(\frac{\Delta - \Delta_{34}}{2}\right)_{r} \left(\frac{\Delta  +\Delta_{34}}{2} \right)_{r+q}}{r! q! \, (\Delta)_{2r+q} \, (\Delta-\mu)_{r} }   
(z \bar z)^r \, (z + \bar z - z \bar z)^q
\eea
To establish this we first note the following identities/definitions:
\bea
(z \bar z)^{\frac{l}{2}} C^\mu_l(\tfrac{z+ \bar z}{2 \sqrt{z \bar z}}) := \sum_{k=0}^{[l/2]} (-1)^k \frac{\left( \mu \right)_{l-k}}{k! \, (l-2k)!} (z+ \bar z)^{l-2k} \, (z \bar z)^{k}
\eea
\bea
(z + \bar z - z \bar z)^q =  \sum_{p=0}^q (-1)^p \left( {q \atop p} \right) \, (z+ \bar z)^{q-p} (z \bar z)^{p}
\eea
Using the double sum identity:
\bea
\sum_{q=0}^\infty \sum_{p=0}^q a_{p, q-p} = \sum_{m=0}^\infty \sum_{n=0}^\infty a_{n,m} = \sum_{l=0}^\infty \sum_{k=0}^{[l/2]} a_{k, l-2k}
\eea
the first expression can be written as
\bea
\sum_{s=0}^\infty \tfrac{\left(\frac{\Delta - \Delta_{12}}{2} - \mu\right)_{s} \left(\frac{\Delta  +\Delta_{34}}{2} - \mu\right)_{s}}{s! \, (\Delta -\mu)_{s}} \sum_{m,n=0}^\infty \tfrac{\left(\frac{\Delta - \Delta_{12}}{2}\right)_{s+m+2n} \left(\frac{\Delta + \Delta_{34}}{2}\right)_{s+m+2n}}{(\Delta)_{s+m+2n} n! \, m!}   
\tfrac{\left(1+\frac{m+2n}{\mu}\right) \left( \mu \right)_{m+n}}{(\mu+1)_{s+m+2n} }  (-1)^n   \, (z \bar z)^{n+s} \, (z+ \bar z)^{m}\nonumber \\ 
\eea
The second of the expressions can be manipulated to:
\bea
\sum_{r=0}^\infty \tfrac{\left(\frac{\Delta + \Delta_{12}}{2}\right)_{r}  \left(\frac{\Delta - \Delta_{34}}{2}\right)_{r} }{r!  \, (\Delta-\mu)_{r} } \sum_{m,p=0}^\infty \tfrac{ \left(\frac{\Delta - \Delta_{12}}{2} \right)_{r+m+p}  \left(\frac{\Delta  +\Delta_{34}}{2} \right)_{r+m+p}}{m! \, p! \, (\Delta)_{2r+m+p}} (-1)^p  (z \bar z)^{r+p} \, (z+\bar z)^{m}
\eea
In the next step we extract the coefficients of $(z \bar z)^q (z+\bar z)^m$ in both these expressions. For this in the first expression we change $n\rightarrow p, ~ s\rightarrow q-p$ and in the second we change $p\rightarrow p, ~ r \rightarrow q-p$. Then in both the expressions the indices $q$ and $m$ run freely over all non-negative integers and the index $p$ runs over $0, 1, \cdots, q$. The corresponding coefficient for the first expression is:
\bea
\label{conjlhs}
\sum_{p=0}^q \tfrac{\left(\frac{\Delta - \Delta_{12}}{2} - \mu\right)_{q-p} \left(\frac{\Delta  +\Delta_{34}}{2} - \mu\right)_{q-p}}{(q-p)! \, (\Delta -\mu)_{q-p}} \tfrac{\left(\frac{\Delta - \Delta_{12}}{2}\right)_{m+p+q} \left(\frac{\Delta + \Delta_{34}}{2}\right)_{m+p+q}}{(\Delta)_{m+p+q} p! \, m!}   
\tfrac{\mu+ m+2p}{(\mu+m+p)_{q+1} }  (-1)^p
\eea
and for the second expression is:
\bea
\label{conjrhs}
\sum_{p=0}^q \tfrac{\left(\frac{\Delta + \Delta_{12}}{2}\right)_{q-p}  \left(\frac{\Delta - \Delta_{34}}{2}\right)_{q-p} }{(q-p)!  \, (\Delta-\mu)_{q-p} }  \tfrac{ \left(\frac{\Delta - \Delta_{12}}{2} \right)_{m+q}  \left(\frac{\Delta  +\Delta_{34}}{2} \right)_{m+q}}{m! \, p! \, (\Delta)_{m+2q-p}} (-1)^p  
\eea
Now the final step is to compare these two expressions (\ref{conjlhs}) and (\ref{conjrhs}) for arbitrary integers $\{d \ge 1, q \ge 0, m \ge 0\}$. We conjecture that these expressions are identical. We have verified this claim for various special cases exactly, and for large subsets of the integer parameters $\{d \ge 1, q \ge 0, m \ge 0\}$ using Mathematica. 
\section{Details of CPW computation in $d=4$}
\label{appC}
To establish the result for four-point scalar CPW in $d=4$ we start by expanding the answer in power series.
\bea
z \, {}_2F_1(\alpha,\beta,\Delta, z) \, {}_2F_1(\alpha -1,\beta-1,\Delta-2, \bar z) &=& \sum_{m=0}^\infty \sum_{n=0}^\infty \frac{\Gamma(\alpha+m) \Gamma(\alpha+n-1) }{\Gamma(\alpha)\Gamma(\alpha-1)} \frac{\Gamma(\beta+m)\Gamma(\beta+n-1)}{\Gamma(\beta)\Gamma(\beta-1)}\cr 
&&\frac{\Gamma(\Delta) \Gamma(\Delta-2)}{\Gamma(\Delta +m)\Gamma(\Delta+n-2)} \frac{z^{m+1} {\bar z}^n}{m!n!} 
\eea
We now divide the {\it rhs} into three terms with $m+1>n$, $m+1<n$ and $m+1=n$. The piece coming from terms with $m+1 =n$ are real and therefore cancel with the corresponding terms from the complex conjugate combination. The remaining parts are obtained by considering the restricted sums
\bea
\sum_{n=0}^\infty \sum_{m=n}^\infty + \sum_{m=0}^\infty \sum_{n=m+2}^\infty 
\eea
Let us consider the conjugate term next:
\bea
\bar z \, {}_2F_1(\alpha,\beta,\Delta, \bar z) \, {}_2F_1(\alpha -1,\beta-1,\Delta-2, z) &=& \sum_{m=0}^\infty \sum_{n=0}^\infty \frac{\Gamma(\alpha+m) \Gamma(\alpha+n-1) }{\Gamma(\alpha)\Gamma(\alpha-1)} \frac{\Gamma(\beta+m)\Gamma(\beta+n-1)}{\Gamma(\beta)\Gamma(\beta-1)}\cr 
&&\frac{\Gamma(\Delta) \Gamma(\Delta-2)}{\Gamma(\Delta +m)\Gamma(\Delta+n-2)} \frac{\bar z^{m+1} {z}^n}{m!n!} 
\eea
again we split this into three types of terms as above and drop the term that is real. Then we can split the rest into two types of terms by writing the sum as before in two parts:
\bea
\sum_{n=0}^\infty \sum_{m=n}^\infty + \sum_{m=0}^\infty \sum_{n=m+2}^\infty 
\eea
Noticing that the first sum in the first term and the second sum in the second have more $z$'s than $\bar z$'s we would like to combine them. In these two we introduce two new variables $m=n+p$ and $n=m+2+q$ to replace $m$ and $n$ respectively. Combing these we have:
\bea
&&  \sum_{n=0}^\infty \sum_{p=0}^\infty\tfrac{\Gamma(\alpha+n+p) \Gamma(\alpha+n-1) }{\Gamma(\alpha)\Gamma(\alpha-1)} \tfrac{\Gamma(\beta+n+p)\Gamma(\beta+n-1)}{\Gamma(\beta)\Gamma(\beta-1)}
\tfrac{\Gamma(\Delta) \Gamma(\Delta-2)}{\Gamma(\Delta +n+p)\Gamma(\Delta+n-2)} \tfrac{z^{n+p+1} {\bar z}^n}{(n+p)!n!} \cr
&&-\sum_{m=0}^\infty \sum_{q=0}^\infty \tfrac{\Gamma(\alpha+m) \Gamma(\alpha+q+m+1) }{\Gamma(\alpha)\Gamma(\alpha-1)} \tfrac{\Gamma(\beta+m)\Gamma(\beta+q+m+1)}{\Gamma(\beta)\Gamma(\beta-1)} \tfrac{\Gamma(\Delta) \Gamma(\Delta-2)}{\Gamma(\Delta +m)\Gamma(\Delta+q+m)} \tfrac{\bar z^{m+1} {z}^{q+m+2}}{m!(q+m+2)!}
\eea
In the second term we can replace $m \rightarrow m-1$ and still sum over the new $m$ from $0$ to $\infty$ as there will be term $(m-1)!$ in the denominator which kills the $m=0$ term. Then
\bea
&&  \sum_{n=0}^\infty \sum_{p=0}^\infty\tfrac{\Gamma(\alpha+n+p) \Gamma(\alpha+n-1) }{\Gamma(\alpha)\Gamma(\alpha-1)} \tfrac{\Gamma(\beta+n+p)\Gamma(\beta+n-1)}{\Gamma(\beta)\Gamma(\beta-1)}
\tfrac{\Gamma(\Delta) \Gamma(\Delta-2)}{\Gamma(\Delta +n+p)\Gamma(\Delta+n-2)} \tfrac{z^{n+p+1} {\bar z}^n}{(n+p)!n!} \cr
&&-\sum_{m=0}^\infty \sum_{q=0}^\infty \tfrac{\Gamma(\alpha+m-1) \Gamma(\alpha+q+m) }{\Gamma(\alpha)\Gamma(\alpha-1)} \tfrac{\Gamma(\beta+m-1)\Gamma(\beta+q+m)}{\Gamma(\beta)\Gamma(\beta-1)} \tfrac{\Gamma(\Delta) \Gamma(\Delta-2)}{\Gamma(\Delta +m-1)\Gamma(\Delta+q+m-1)} \tfrac{\bar z^{m} {z}^{q+m+1}}{(m-1)!(q+m+1)!}
\eea
Now we change dummy variables $n \rightarrow s$, $p\rightarrow l$ in the first term and $m \rightarrow s$ and $q \rightarrow l$ in the second term and combine terms to write this as:
\bea
&& \sum_{l=0}^\infty \sum_{s=0}^\infty\tfrac{\Gamma(\alpha+l+s) \Gamma(\alpha+s-1) }{\Gamma(\alpha)\Gamma(\alpha-1)} \tfrac{\Gamma(\beta+l+s)\Gamma(\beta+s-1)}{\Gamma(\beta)\Gamma(\beta-1)} \tfrac{\Gamma(\Delta) \Gamma(\Delta-2)}{\Gamma(\Delta+s-2)\Gamma(\Delta+l+s-1)} \tfrac{z^{l+s+1} {\bar z}^s}{(s-1)! (l+s)!}\cr
&& \left[
\tfrac{1}{(\Delta +l+s-1) \, s}  -\tfrac{1}{(\Delta +s-2)(l+s+1)}\right]
\eea
Using
\bea
\tfrac{1}{(\Delta +l+s-1) \, s}  -\tfrac{1}{(\Delta +s-2)\, (l+s+1)} = \tfrac{(\Delta-2)(l+1)}{(\Delta +l+s-1)(\Delta +s-2)(l+s+1) s}
\eea
This can be seen to be:
\bea
&& \sum_{l=0}^\infty \sum_{s=0}^\infty\tfrac{\Gamma(\alpha+l+s) \Gamma(\alpha+s-1) }{\Gamma(\alpha)\Gamma(\alpha-1)} \tfrac{\Gamma(\beta+l+s)\Gamma(\beta+s-1)}{\Gamma(\beta)\Gamma(\beta-1)} \tfrac{\Gamma(\Delta) \Gamma(\Delta-1)}{\Gamma(\Delta+s-1)\Gamma(\Delta+l+s)} \tfrac{z^{l+s+1} {\bar z}^s}{s! (l+s+1)!}\eea
which is precisely the first term in our OWN computation of the block. The remaining two terms are simply conjugates of what we have dealt with so far and therefore are going to reproduce the second term in our OWN computation.
%
%
%
%
\bibliographystyle{utphys}
\providecommand{\href}[2]{#2}\begingroup\raggedright\endgroup

\end{document}